\documentclass[prc,aps,floats,floatfix,twocolumn,showpacs,nofootinbib,superscriptaddress]{revtex4}

\usepackage{amsmath}
\usepackage{amssymb}
\usepackage{amsfonts}
\usepackage{graphicx}
\usepackage{tabularx}
\usepackage{multirow}
\usepackage{lscape}
\usepackage{rotating}

\usepackage{pstricks}

\renewcommand{\vec}[1]{{\mathbf #1}}

\newcommand{\nn}{\nonumber}
\newcommand{\noi}{\noindent}

\newcommand{\vsigma}{\boldsymbol{\sigma}}
\newcommand{\bbK}{\boldsymbol{\mathbb K}}

\newcommand{\bbKzz}{\bbK_{\rm  0,0}}
\newcommand{\bbKzu}{\bbK_{\rm  0,1}}
\newcommand{\bbKuz}{\bbK_{\rm  1,0}}
\newcommand{\bbKuu}{\bbK_{\rm  1,1}}
\newcommand{\bbKmz}{\bbK_{\rm -1,0}}
\newcommand{\bbKum}{\bbK_{\rm  1,-1}}
\newcommand{\bbKmm}{\bbK_{\rm -1,-1}}

\newcommand{\wh}{\widehat}

\newcommand{\alp}{\alpha}

\newcommand{\wzu}{W^{(0)}_1}
\newcommand{\wuu}{W^{(1)}_1}
\newcommand{\wzd}{W^{(0)}_2}
\newcommand{\wud}{W^{(1)}_2}

\newcommand{\whuzu}{{\wh W}^{(1,0)}_1}
\newcommand{\whuuu}{{\wh W}^{(1,\pm 1)}_1}

\newcommand{\pbdbt}{\left( \beta_2 - \beta_3 \right)}

\newcommand{\chiz}{\chi_0}
\newcommand{\chid}{\chi_2}
\newcommand{\chiq}{\chi_4}

\newcommand{\pdemi}{\tfrac{1}{2}}

\newcommand{\moq}{\frac{m^*\omega}{q}}

\newcommand{\moqd}{\left( \moq \right)^2}

\newcommand{\mrho}{m^* \rho}

\newcommand{\bc}{\begin{center}}
\newcommand{\ec}{\end{center}}
\newcommand{\be}{\begin{equation}}
\newcommand{\ee}{\end{equation}}
\newcommand{\bqr}{\begin{eqnarray}}
\newcommand{\eqr}{\end{eqnarray}}
\newcommand{\bi}{\begin{itemize}}
\newcommand{\ei}{\end{itemize}}
\newcommand{\bwt}{\begin{widetext}}
\newcommand{\ewt}{\end{widetext}}

\newcommand{\citeqdot}[1]{Eq.~(\ref{#1})}

\newcommand{\citeqssdot}[3]{Eqs.~(\ref{#1}), (\ref{#2}) and~(\ref{#3})}

\newcommand{\citerefsdot}[1]{refs.\cite{#1}}

\newcommand{\citeFigure}[1]{Fig.~\ref{#1}}

\newcommand{\citeAppendix}[1]{Appendix~\ref{#1}}

\begin{document}

\title{Nuclear response for the Skyrme effective interaction with
       zero-range tensor terms. III. Neutron matter and neutrino propagation}


\author{A. Pastore}
\email{pastore@ipnl.in2p3.fr}
\affiliation{Universit\'e de Lyon, F-69003 Lyon, France; Universit\'e Lyon 1,
             43 Bd. du 11 Novembre 1918, F-69622 Villeurbanne cedex, France\\
             CNRS-IN2P3, UMR 5822, Institut de Physique Nucl{\'e}aire de Lyon}

\author{M. Martini}
\email{mmartini@ulb.ac.be}
\affiliation{Institut d'Astronomie et d'Astrophysique, CP-226,
             Universit{\'e} Libre de Bruxelles, 1050 Brussels, Belgium}

\author{V. Buridon}
\email{v.buridon@ipnl.in2p3.fr}
\affiliation{Universit\'e de Lyon, F-69003 Lyon, France; Universit\'e Lyon 1,
             43 Bd. du 11 Novembre 1918, F-69622 Villeurbanne cedex, France\\
             CNRS-IN2P3, UMR 5822, Institut de Physique Nucl{\'e}aire de Lyon}

\author{D. Davesne}
\email{davesne@ipnl.in2p3.fr}
\affiliation{Universit\'e de Lyon, F-69003 Lyon, France; Universit\'e Lyon 1,
             43 Bd. du 11 Novembre 1918, F-69622 Villeurbanne cedex, France\\
             CNRS-IN2P3, UMR 5822, Institut de Physique Nucl{\'e}aire de Lyon}
             
\author{K. Bennaceur}
\email{bennaceur@ipnl.in2p3.fr}
\affiliation{Universit\'e de Lyon, F-69003 Lyon, France; Universit\'e Lyon 1,
             43 Bd. du 11 Novembre 1918, F-69622 Villeurbanne cedex, France\\
             CNRS-IN2P3, UMR 5822, Institut de Physique Nucl{\'e}aire de Lyon}

\author{J. Meyer}
\email{jmeyer@ipnl.in2p3.fr}
\affiliation{Universit\'e de Lyon, F-69003 Lyon, France; Universit\'e Lyon 1,
             43 Bd. du 11 Novembre 1918, F-69622 Villeurbanne cedex, France\\
             CNRS-IN2P3, UMR 5822, Institut de Physique Nucl{\'e}aire de Lyon}


\begin{abstract}
The formalism of the linear response  for the Skyrme energy density
functional including tensor terms derived in
articles~\cite{Davesne09,Davesne12} for nuclear matter 
is applied here to the case of pure neutron matter.  As in
article~\cite{Davesne12} we present analytical results for the response
function in all channels, the Landau parameters and the odd-power sum rules. 
Special emphasis is given to the inverse energy weighted sum rule because it
can be used to detect non physical instabilities.  Typical examples are
discussed and numerical results shown. Moreover, as a direct application,
neutrino propagation in neutron matter is investigated through its neutrino
mean free path at zero temperature. This quantity turns out to be very
sensitive to the tensor terms of the Skyrme energy density functional.  
\end{abstract}


 \date{\today}


\maketitle


\section{Introduction}
\label{sect:intro}


In a recent series of articles~\cite{Davesne09,Davesne12}, hereafter denoted respectively article I and II,
the  contribution of the zero-range tensor terms in the Skyrme effective interaction has been analyzed in the context of the 
 Linear Response theory. The first result from these articles is that the tensor terms have  very sizable effects on the response functions. 
Another important result is that the inverse energy weighted sum rule can be used as a tool of diagnosis for instabilities. These two articles were devoted to Symmetric Nuclear Matter (SNM) only. Since the construction of an Energy Density Functional (EDF) reliable for both symmetric matter and neutron matter is of fundamental importance~\cite{Kortelainen:2010hv,Kortelainen12}, 
we present here the response functions and some associated sum rules for Pure Neutron Matter (PNM) with the same approach.
The interest of the present study is related to spin susceptibilities and ferromagnetic
finite size instabilities in neutron
matter~\cite{Marcos91,Fantoni01,Margueron02,Vidana02a,Vidana02b,
Isayev04,Beraudo04,Beraudo05,Rios05,Bombaci:2005vi,Lopez-Val06,Krastev07,
Bordbar08,Margueron:2009rn,Isayev:2009nt,Chamel:2010wr}. 
Moreover, we use these results to study the impact of the tensor terms on the determination of the neutrino mean free path in PNM. 
This is a quantity of intrinsic importance
since the cooling of a neutron star core in its first moments is governed by neutrino emission and therefore by their mean free path trough dense matter. Some previous studies 
using non-relativistic approaches~\cite{Iwamoto:1982zp,Reddy:1998hb,Navarro:1999jn,Shen:2003ih,Margueron:2003fq,Margueron06} have revealed some very interesting features of the mean free path properties but they all neglected the
possible tensor contribution. Since the neutrino mean free path is directly related to the response functions which are themselves affected by the tensor, it is worthwhile to determine precisely the induced modifications. 

The article is organized as follows: in the first part devoted to the linear response theory approach, we present explicit expressions for the spin response functions, the Landau parameters and the sum rules $M_1, M_3,M_{-1}$. Since the technical approach follows closely those of the previous articles, this part mainly contains figures and discussion, formula being written in appendices. The second part deals with the problem of neutrino mean free path. We first give explicit expression of this quantity in presence of tensor interaction then we show the influence of the
parameterizations of the Skyrme functional.


\section{Linear response approach to neutron matter}
\subsection{Response function}
\label{subsec_lr}
Following article II, the starting point for the determination of the response functions is the Skyrme energy functional. Since in neutron matter the isospin is no longer a relevant quantum number and isovector and isoscalar densities are equal, it is convenient to define new coupling constants $C^{x} = C_{0}^{x} + C_{1}^{x}$, where $x=\rho$, $\tau$, $\Delta\rho, ...$  in such a way that the Energy Density Functional  can be written as
\begin{equation}
E=\int{\cal E}\,\mathrm d^3r
\end{equation}
with
\bqr
{\cal E} & = & C^{\rho} \left[ \rho_{} \right] \rho_{}^{2} 
           +   C^{\Delta\rho} \rho_{} \Delta\rho_{} 
           +   C^{\tau}( \rho_{} \tau_{}                                                        
           -   \mathbf{j}_{}^{2} )                                                         \\ 
         & + & C^{s} \left[ \rho_{} \right] \mathbf{s}^{2}_{}
           +   C^{\nabla \mathbf{s}} \left( \nabla \cdot \mathbf{s}_{} \right)^{2}
           +   C^{\Delta s}  \mathbf{s}_{} \cdot \Delta \mathbf{s}_{}                                        \nn  \\
         & + & C^{T} \left(  \mathbf{s}_{} \cdot \mathbf{T}_{} -\sum_{\mu\nu=x}^{z} J_{\mu\nu} J_{\mu\nu}\right) 
                                                      \nn  \\
         & + &     C^{F} \left[ s_{} \cdot \mathbf{F} -\frac{1}{2}\left( \sum_{\mu=x}^{z} J_{\mu\mu} \right)^{2} -\frac{1}{2}\sum_{\mu\nu=x}^{z} J_{\mu\nu} J_{\nu\mu}  \right] \nn\\
         &+& C^{\nabla J} \left[ \rho \nabla \cdot \mathbf{J} 
           +   \mathbf{s} \cdot \left( \nabla \times \mathbf{j} \right)\right].      \nn    
\label{eq:EF:full}
\eqr

The expressions of the coupling constants as functions of the parameters of the Skyrme interaction can be found in article I.
From this expression, it is straightforward to obtain the residual interaction (see table in~\citeAppendix{app:phme_tensor}) by taking the second derivative of the EDF with respect to the density.

The Random Phase Approximation (RPA) response function in each channel $(\alpha)=(\text{S,M})\equiv$ (spin, projection of the spin)
(see~\citeAppendix{app:rf_pnm}) is then obtained by solving the Bethe-Salpeter equations for the correlated Green functions $G_{RPA}^{(\alp)}$. Finally, from these response functions, we can easily obtain the Landau parameters (see~\citeAppendix{app:landau}).
The quantities of interest are not directly the RPA propagators themselves, but merely the response functions
  $S^{(\alp)}(\mathbf{q},\omega)$, also called the dynamical structure functions by some authors,  which are
defined at zero temperature by
\be 
 S^{(\alp)}(\mathbf{q},\omega)
   = -\frac{1}{\pi} \, \rm{Im} \, \chi^{(\alp)}(\mathbf{q},\omega)\,. 
\label{eq:sfunc}
\ee

\begin{figure}[t]
\bc
$\begin{matrix}
     \includegraphics[clip,scale=0.155,angle=-90]{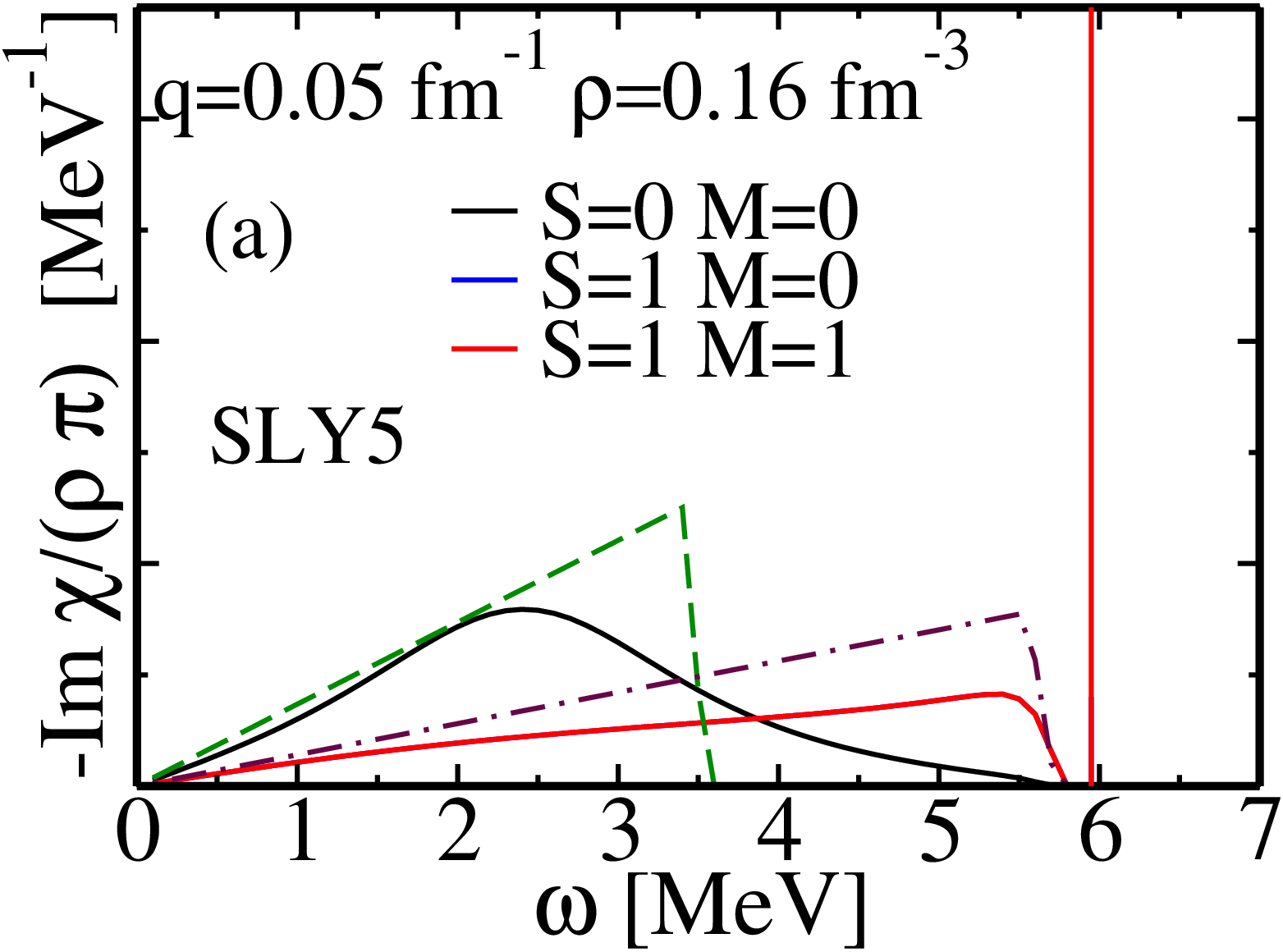} \!\!\!&\!\!\!
     \includegraphics[clip,scale=0.155,angle=-90]{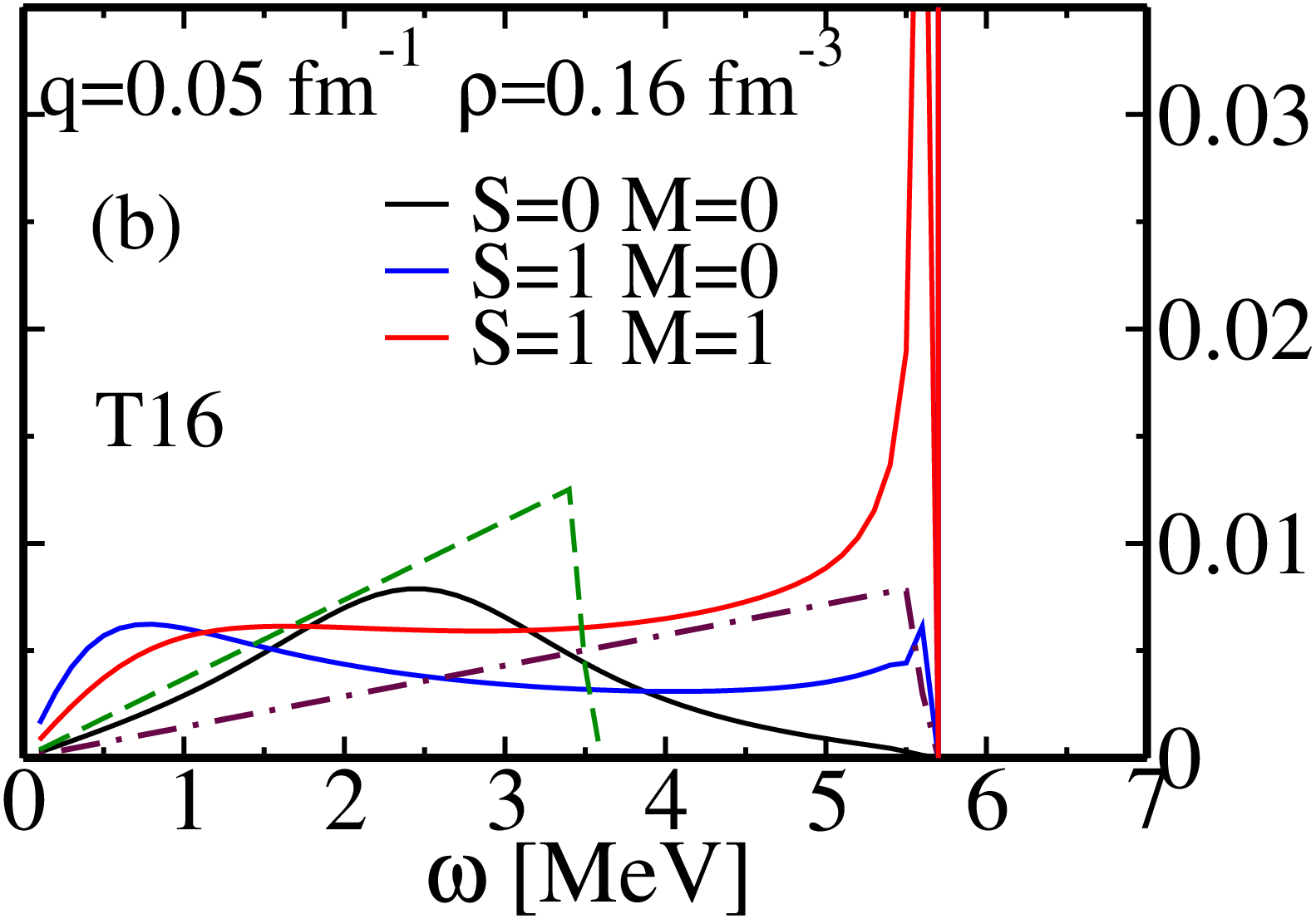} \\
     \includegraphics[clip,scale=0.155,angle=-90]{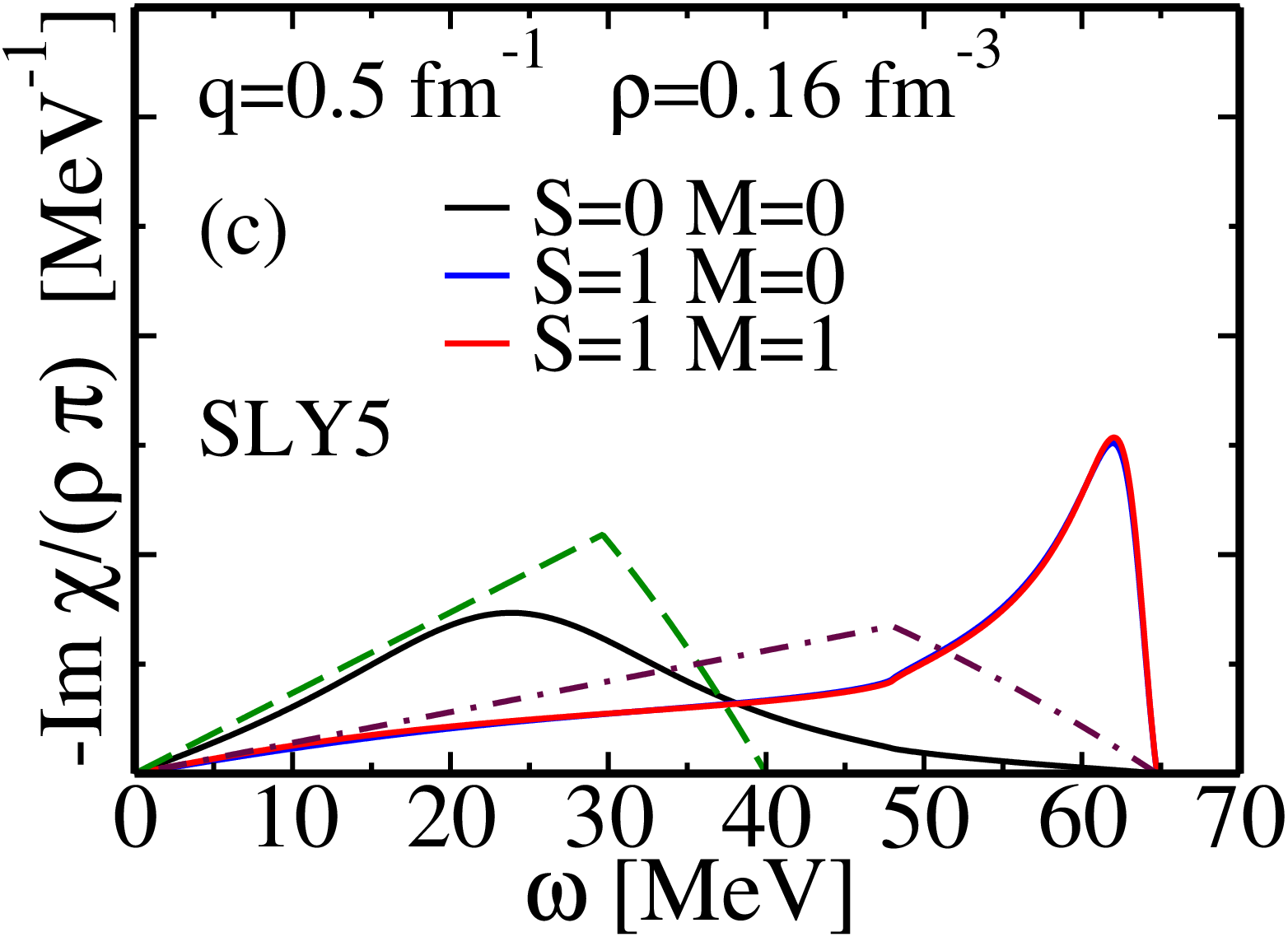} \!\!\!&\!\!\!
     \includegraphics[clip,scale=0.155,angle=-90]{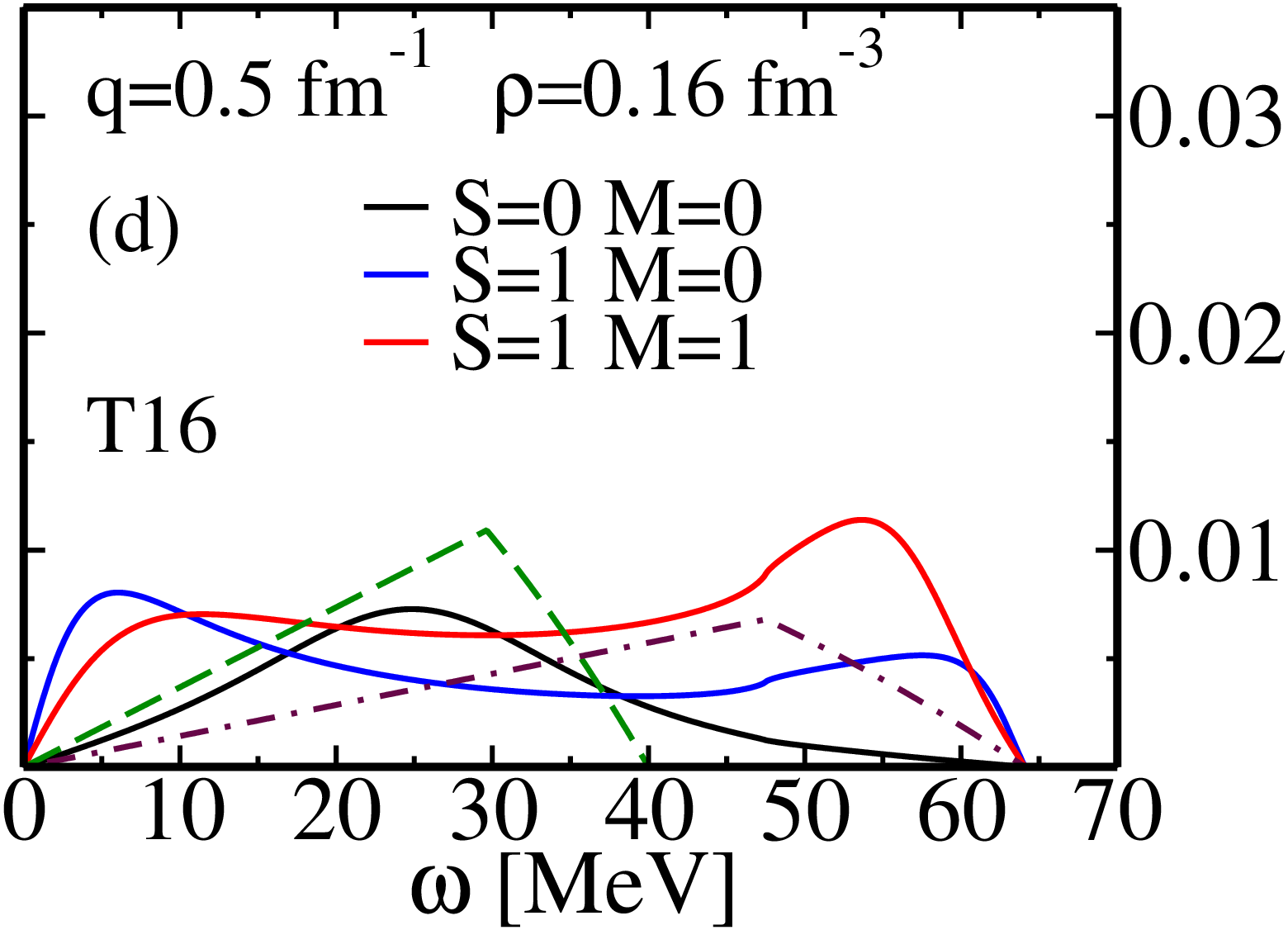} \\
     \includegraphics[clip,scale=0.155,angle=-90]{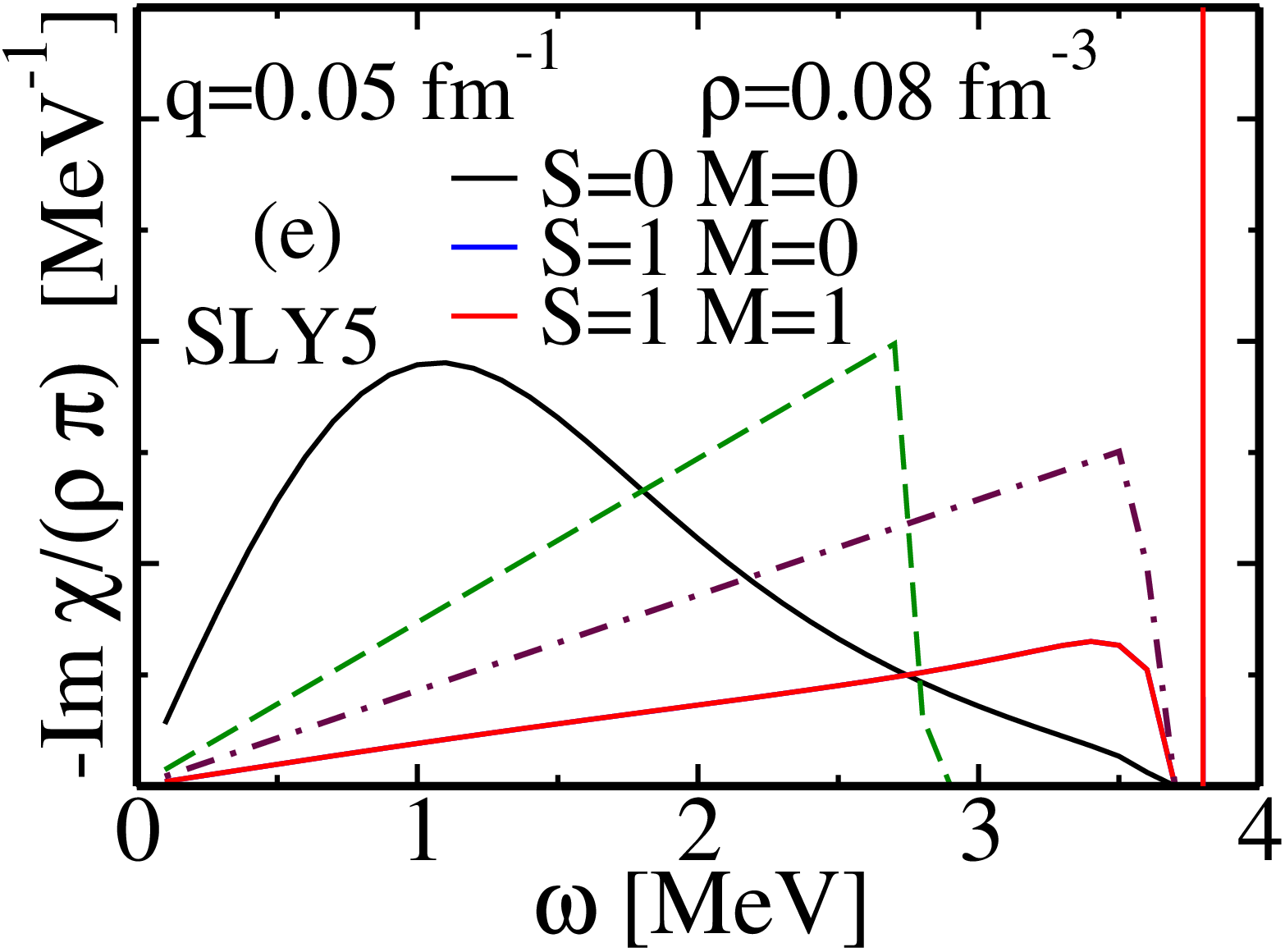} \!\!\!&\!\!\!
     \includegraphics[clip,scale=0.155,angle=-90]{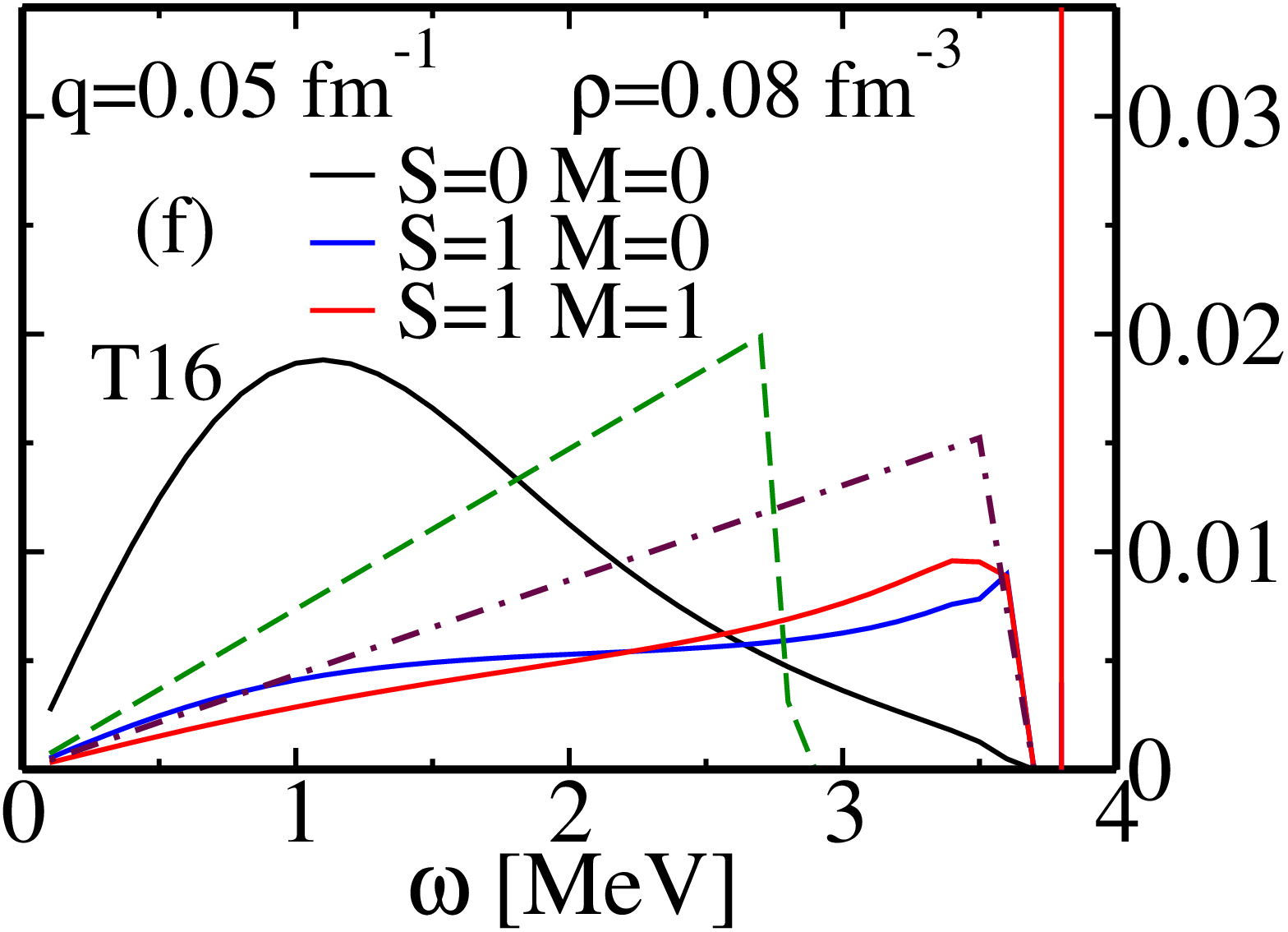} \\
     \includegraphics[clip,scale=0.155,angle=-90]{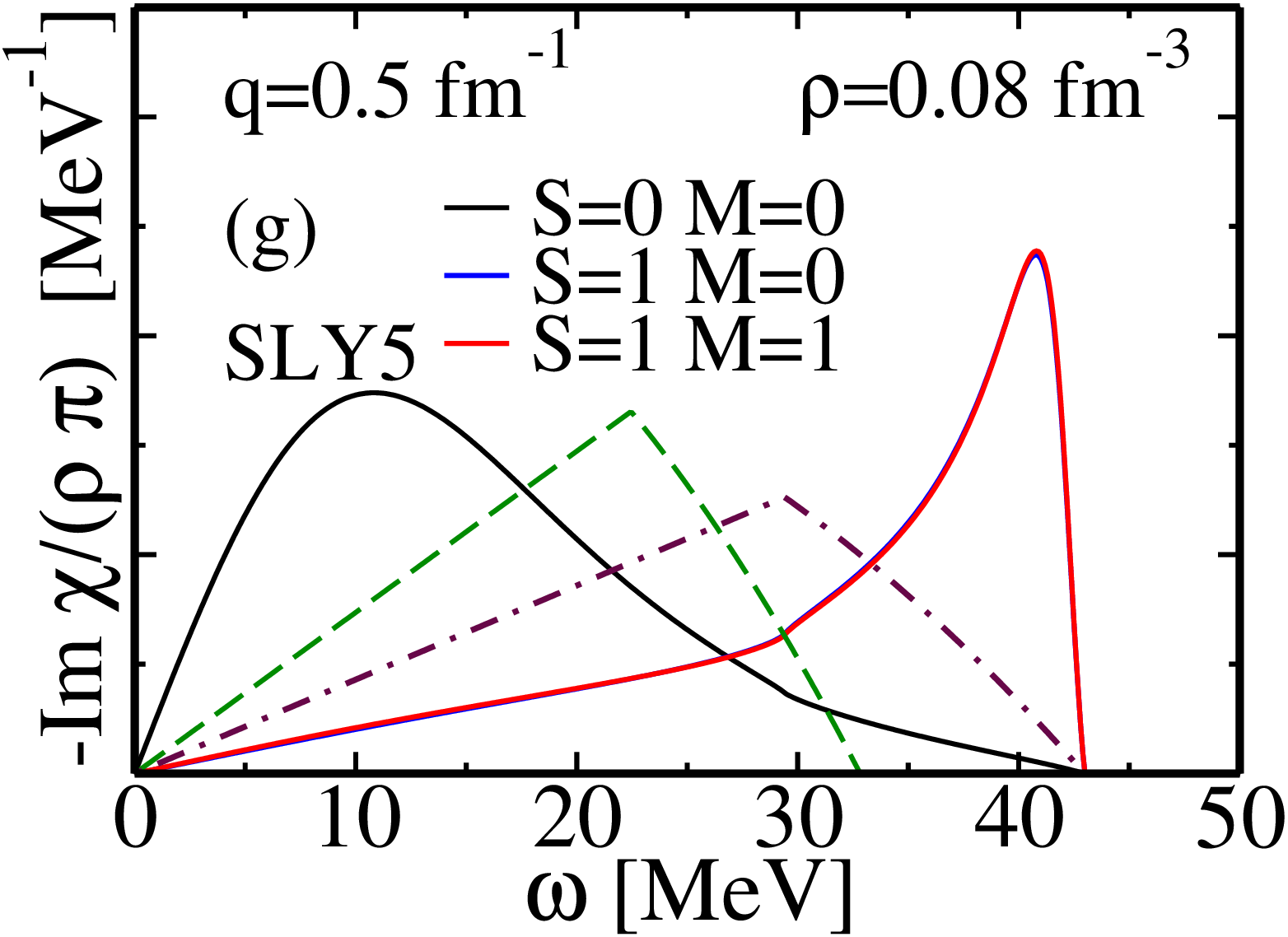} \!\!\!&\!\!\!
     \includegraphics[clip,scale=0.155,angle=-90]{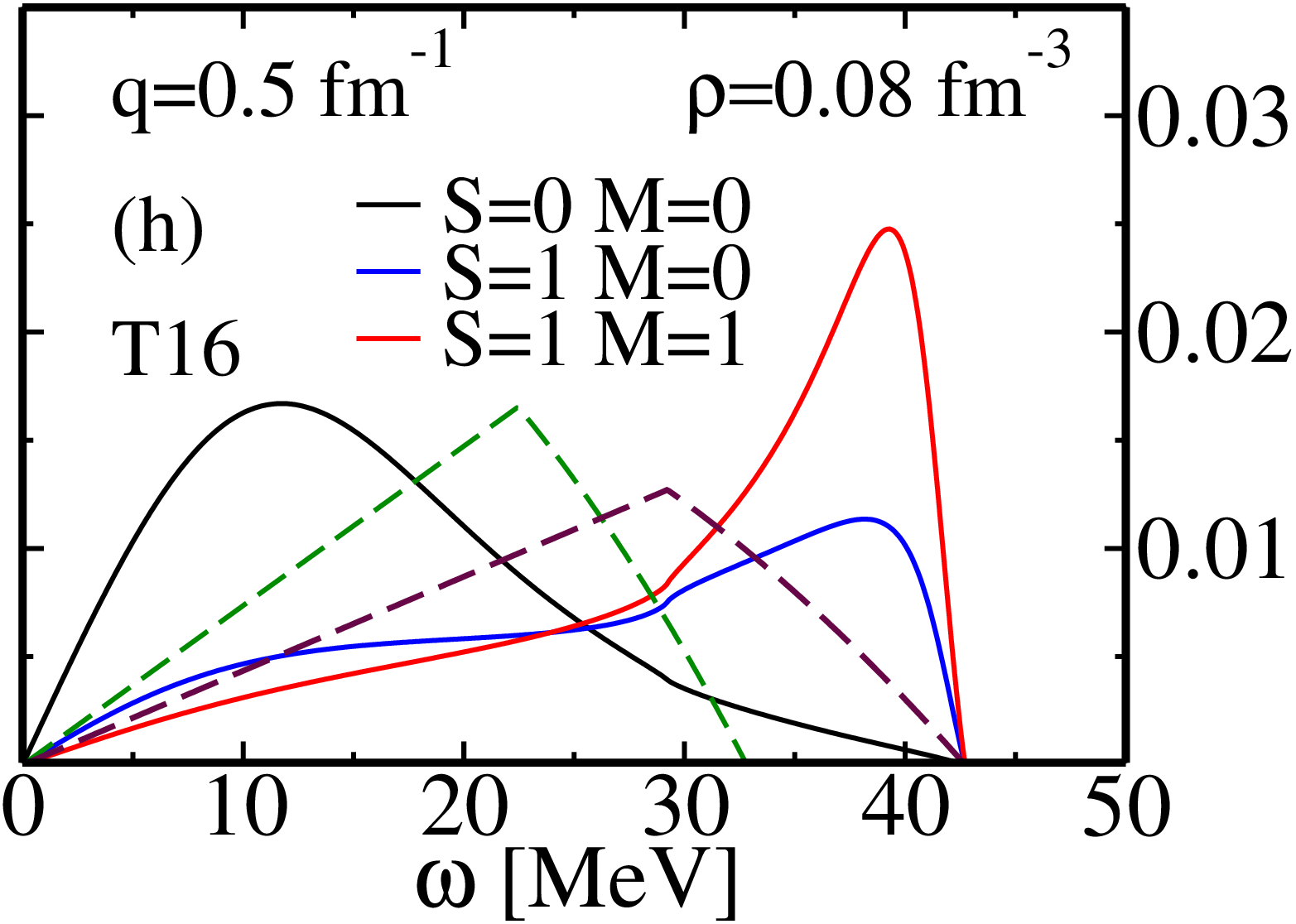} \\
\end{matrix}$
\ec
      \caption{(Color on line) For the three PNM channels, we show the
                response functions $S^{(\alp)} (q,\omega)$ for the
                interaction SLy5 and T16. The vertical lines represent the
                position of the eventual zero-sound mode. For each force we
                plot as a reference the Fermi gas response function (dashed
                line) and the uncorrelated response function (dashed-dotted
                line).}
\label{fig:marco1}
\end{figure}


\begin{figure}[htbp]
      {\includegraphics[width=0.75\columnwidth,angle=-90,clip]{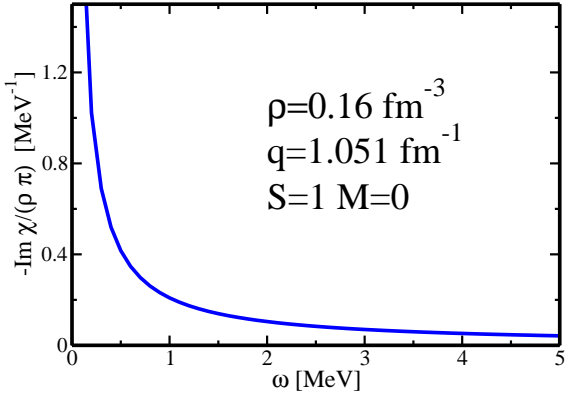}
       \caption{(Colors online) The response functions $S^{(\alp)} (q,\omega)$
                calculated for the Skyrme tensor parameterization
                T16~\cite{Lesinski07}, for the  channel $S=1,\,M=0$ only.
		The transfer momentum is $q=1.051$~fm$^{-1}$ and the density of the system is $\rho = 0.16$~fm$^{-3}$.}
       \label{fig:T16}}
\end{figure}

\begin{figure}[ht]
      \includegraphics[clip,scale=0.345,angle=-90,bb=80 0 588 708]{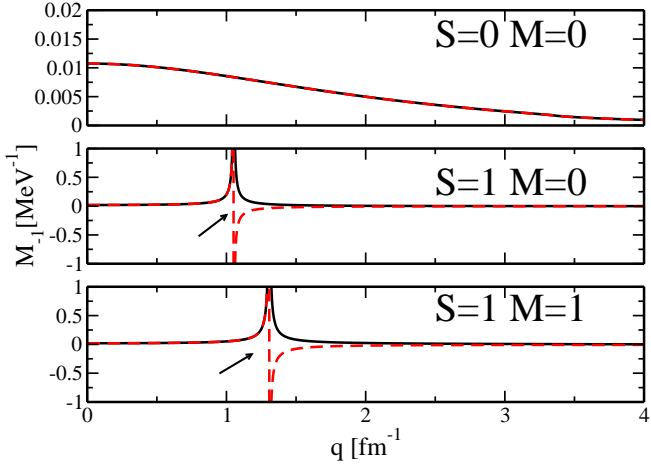}
      \caption{(Color on line) IEWSR (in MeV$^{-1}$) as function of the transferred momentum $q$ (in fm$^{-1}$) for the T16 tensor parameterization. Full black line shows the result of the integral~\citeqdot{integral} while the dashed red line shows the result of the analytical expression~\citeqssdot{eq:m_-1:00}{eq:m_-1:10}{eq:m_-1:11}. The results are obtained at  $\rho=0.16$~fm$^{-3}$. The arrow indicates the position of the instability.} 
\label{fig:m-1_T16}
\end{figure}
From now on we choose the direction of $\mathbf{q}$ along the $z$ axis, as done in article I and II. 
We show this function in Fig.~\ref{fig:marco1} for two different values of
the transferred momentum 
($q=0.05$~fm$^{-1}$ and $q=0.5$~fm$^{-1}$ and two different densities
($\rho=0.08~\mathrm{fm}^{-3}$ and $\rho=0.16$~fm$^{-3}$) as a function of the energy $\omega$. 
As in article I and II we use a system of natural units so that  $\hbar=c=1$.
We consider one interaction without tensor, {\em i.e.} SLy5, and one with tensor, T16 (see article of Lesinski \emph{et al.}~\cite{Lesinski07} for the definition of the TIJ parameterizations). 
Among the several TIJ possibilities the choice of T16 is motivated by the study
of the neutrino mean free path (see later).  
In order to illustrate the effect of the interaction and of the RPA correlations we plot in each panel
the corresponding Fermi gas (FG) and Hartree-Fock results ({\em i.e.} uncorrelated response functions). 
A first effect of the interaction clearly appears at the Hartree-Fock level where the mean field is  responsible for the dressing of the bare neutron mass giving a density-dependent effective one. The difference between the Fermi gas and Hartree-Fock structure functions increases with the density.   
With the RPA correlations, the difference between the $S=0$ and $S=1$ channel of the p-h interaction are reflected in the corresponding response functions. Let us focus on the S=1 channel, particularly important for the neutrino mean free path. The ($S=1,\,M=0$) and ($S=1,\,M=1$) structure functions practically coincide, 
as it is expected for the SLy5 force, while they clearly differ in the presence of tensor interaction. 
Concerning the $S=1$ for low $q$ (as illustrated in the figure for $q=0.05$~fm$^{-1}$) a spin zero-sound mode appears. It stands out above the p-h continuum. The existence of this spin collective mode, called magnon, makes harder the excitation of the system, hence, correspondingly the p-h continuum response is depleted. This anti-ferromagnetic behaviour disappears when $q$ increases. For high $q$ another kind of divergence may appear. As illustrated in
Fig.~\ref{fig:T16}, the enhancement 
of the response function may become dramatic and show a pole at $\omega=0$.
In this case the homogeneous Hartree-Fock ground state becomes unstable. For lower values of $q$ the same kind of instability appears at higher density, called critical density $\rho_c$, as illustrated in the next section.

\subsection{Sum rules and moments of the strength function}
%

\begin{figure}[t]
\bc
$\begin{matrix}
      \includegraphics[clip,scale=0.165,angle=-90,bb=0 0 612 730]{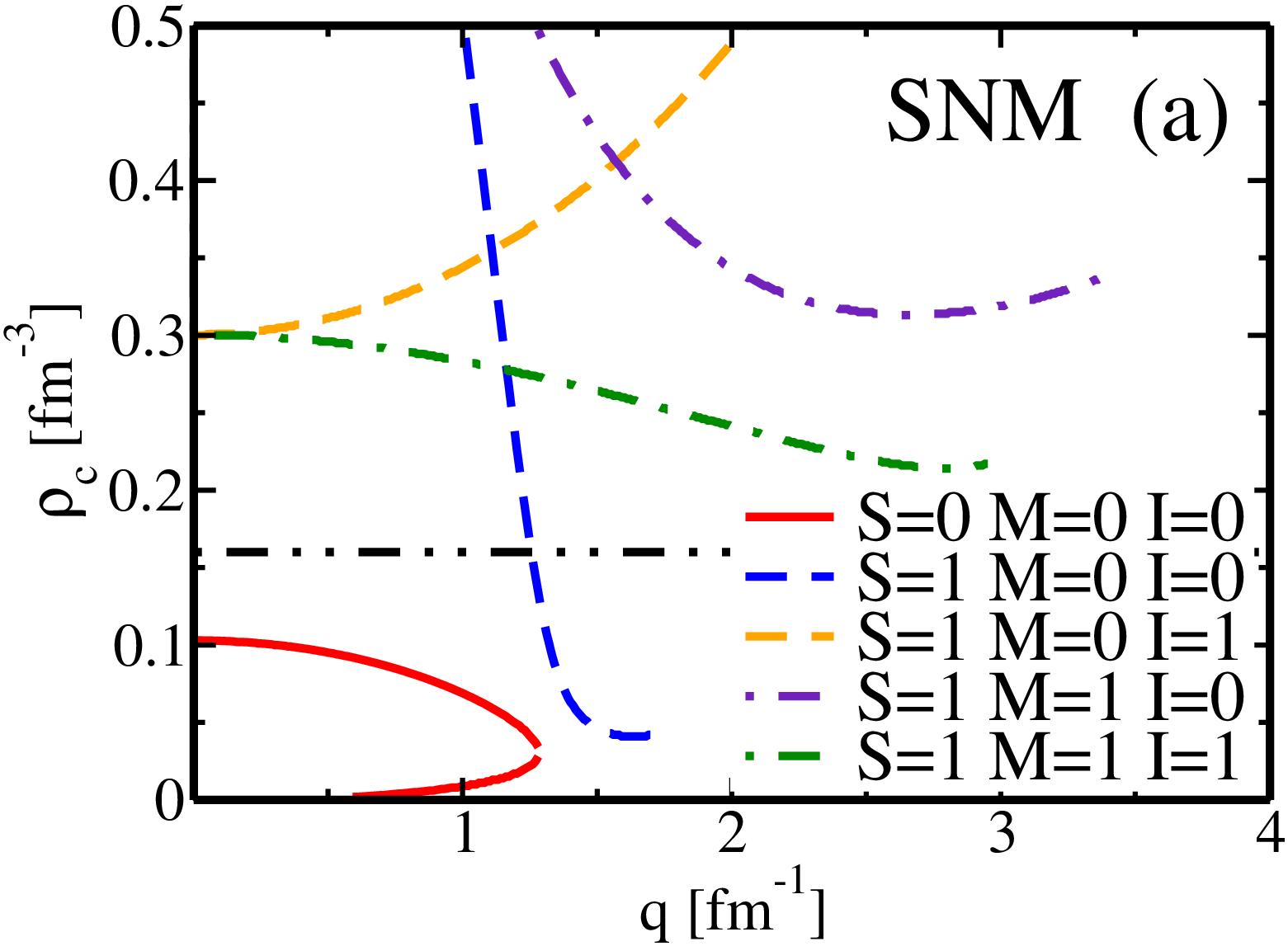} &
      \includegraphics[clip,scale=0.165,angle=-90,bb=0 0 612 730]{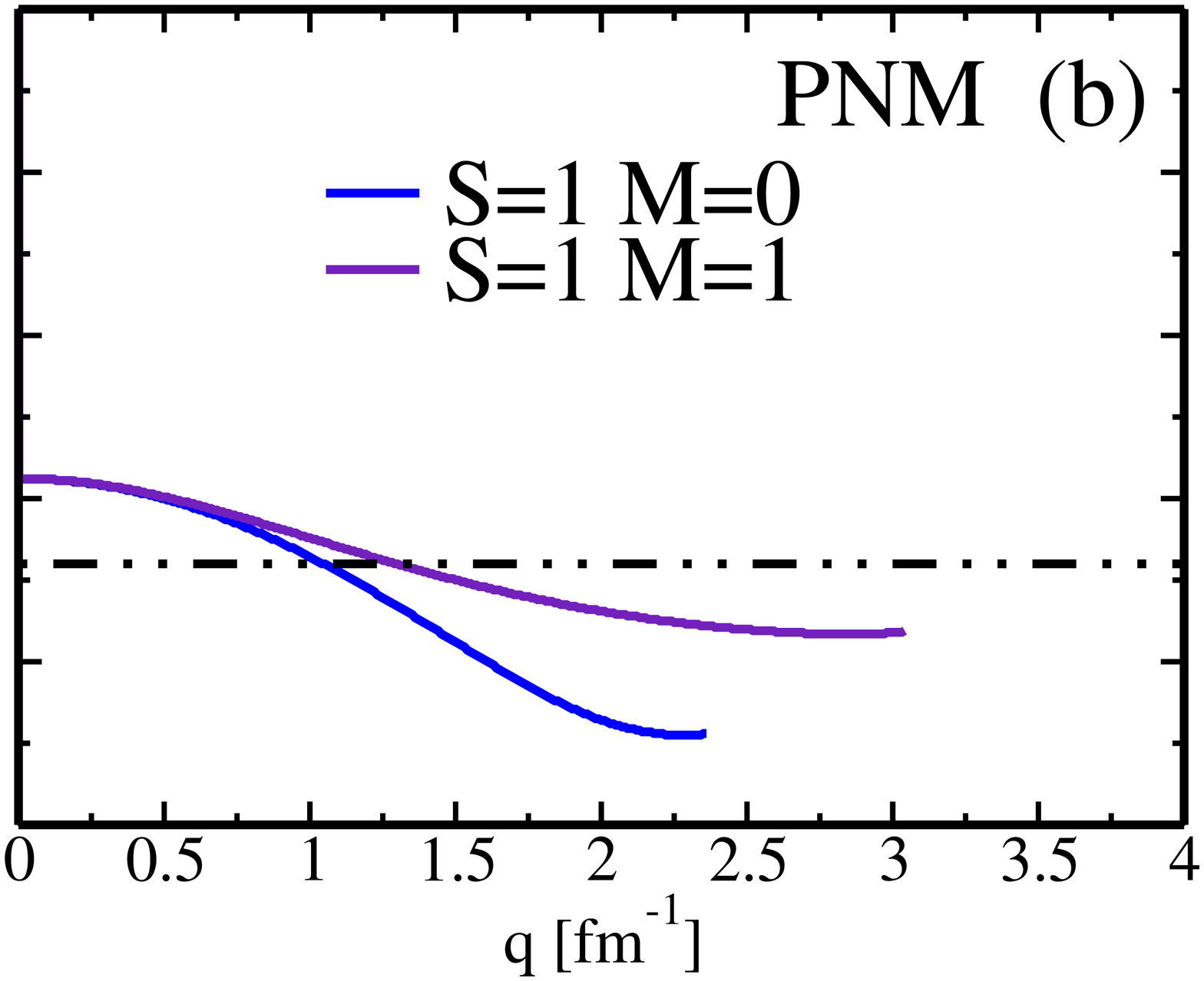}
\end{matrix}$
\ec
      \caption{(Color on line) Critical densities (in fm$^{-3}$) as 
               functions of the transferred momentum $q$ (in fm$^{-1}$) 
               for the T16 tensor parameterization and for 
               SNM (panel (a)) and for PNM (panel (b)). The horizontal dashed-dotted line is placed at $\rho=0.16$~fm$^{-3}$ just to guide the eye.}
\label{fig:critical_T16}
\end{figure}

\begin{figure}[htb]
\bc
$\begin{matrix}
    \includegraphics[clip,scale=0.16,angle=-90,bb=0 0 612 725]{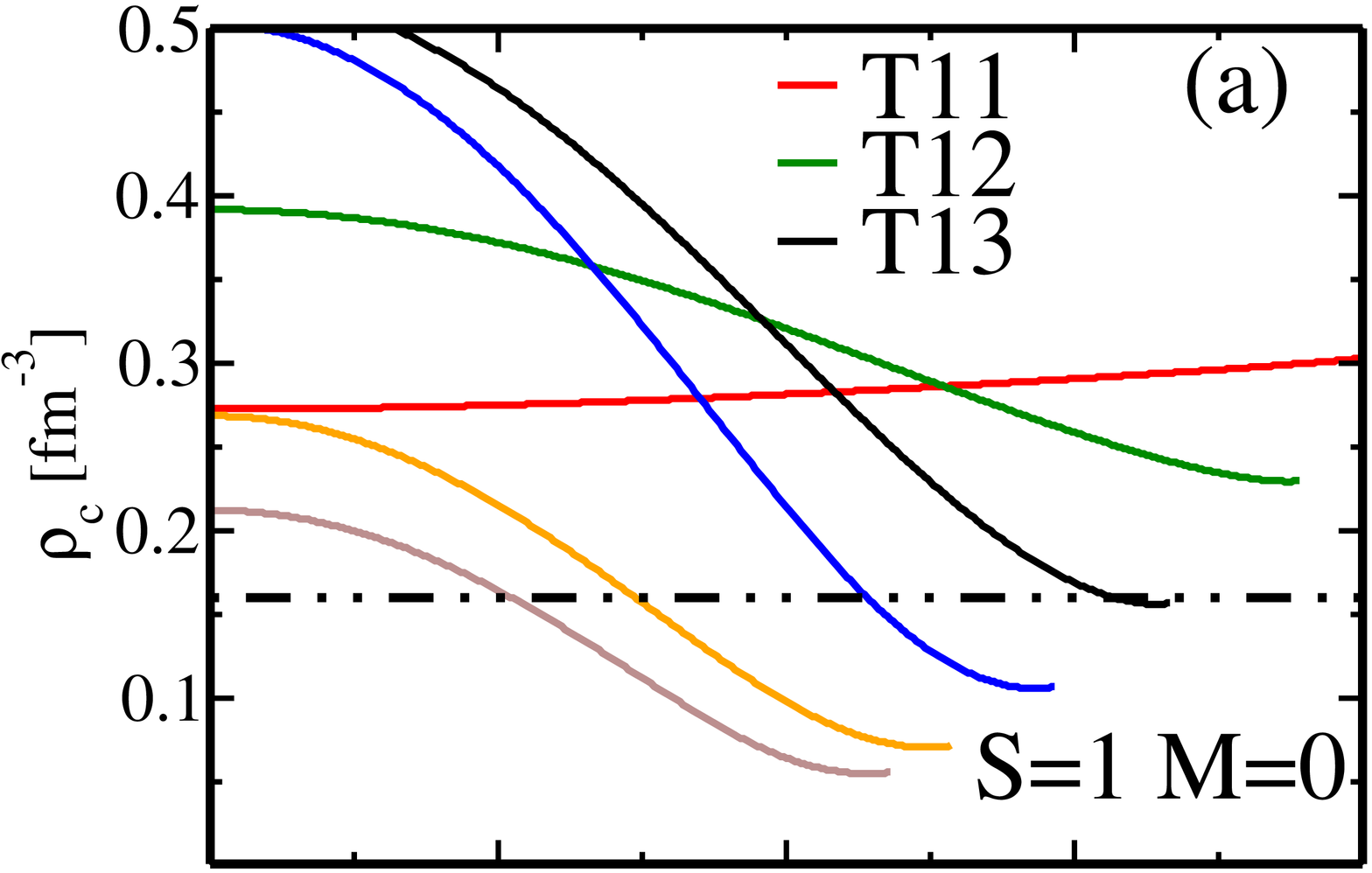} &
    \includegraphics[clip,scale=0.16,angle=-90,bb=0 0 612 725]{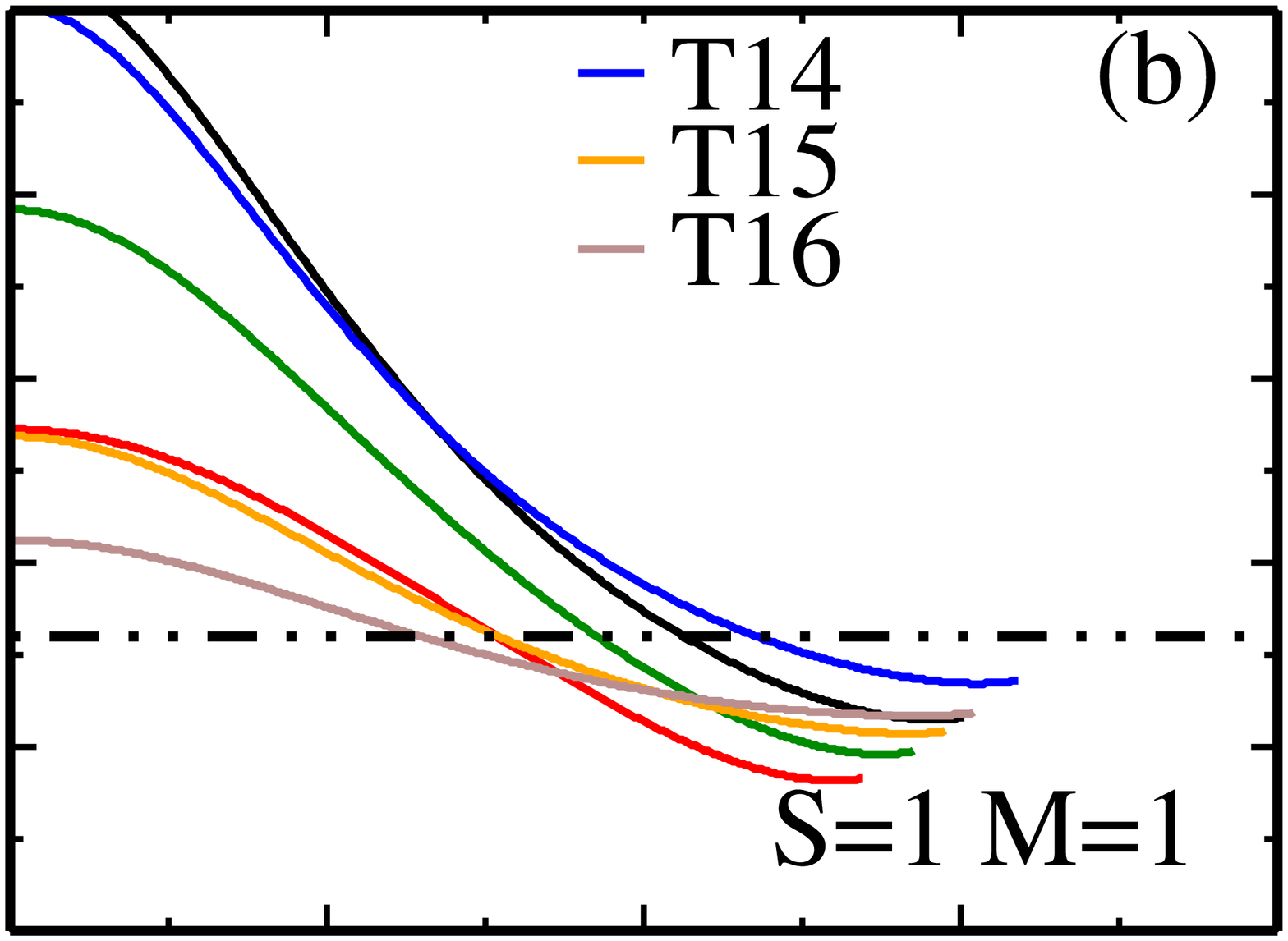} \\
    \includegraphics[clip,scale=0.16,angle=-90,bb=0 0 612 725]{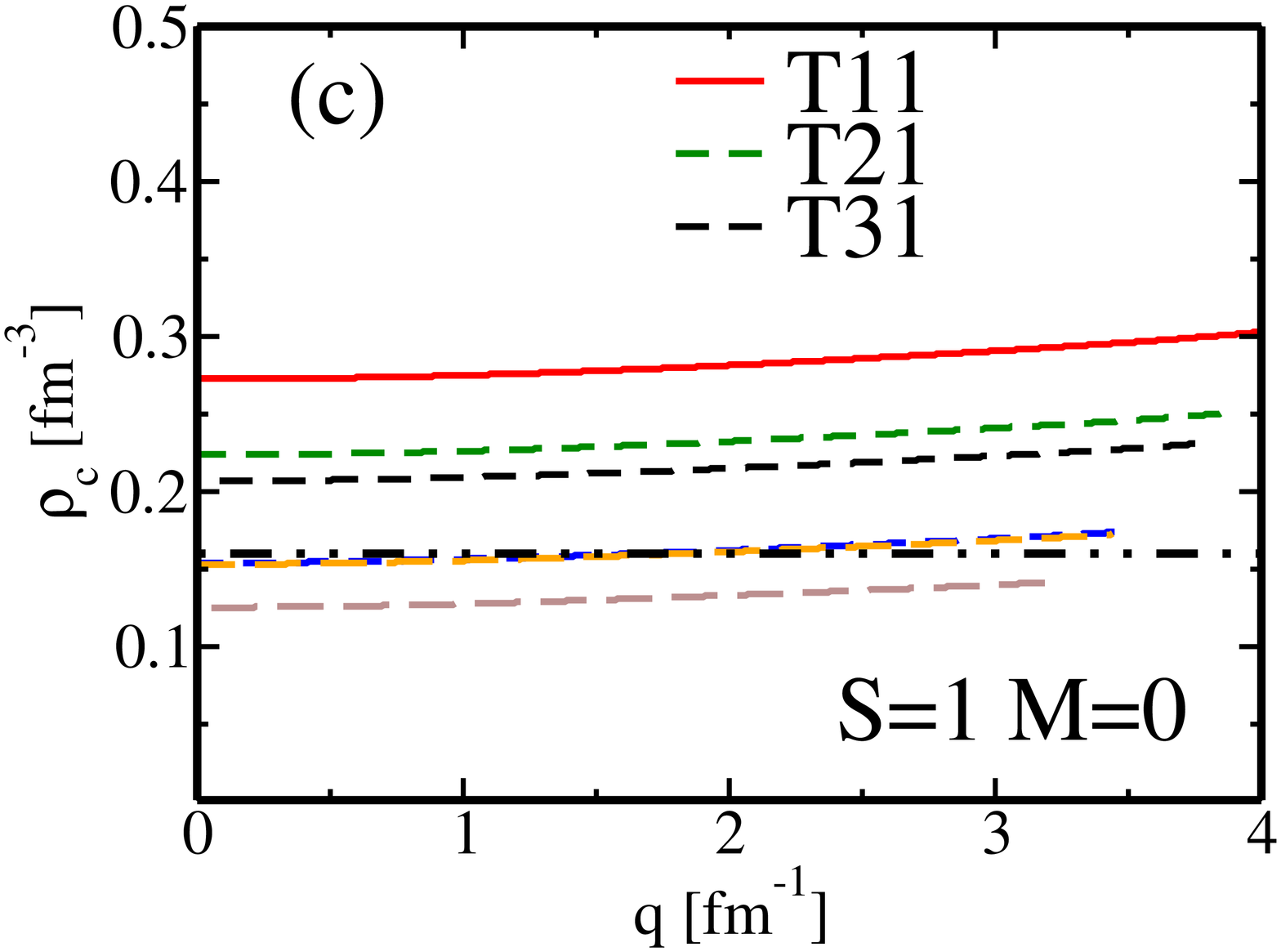} &
    \includegraphics[clip,scale=0.16,angle=-90,bb=0 0 612 725]{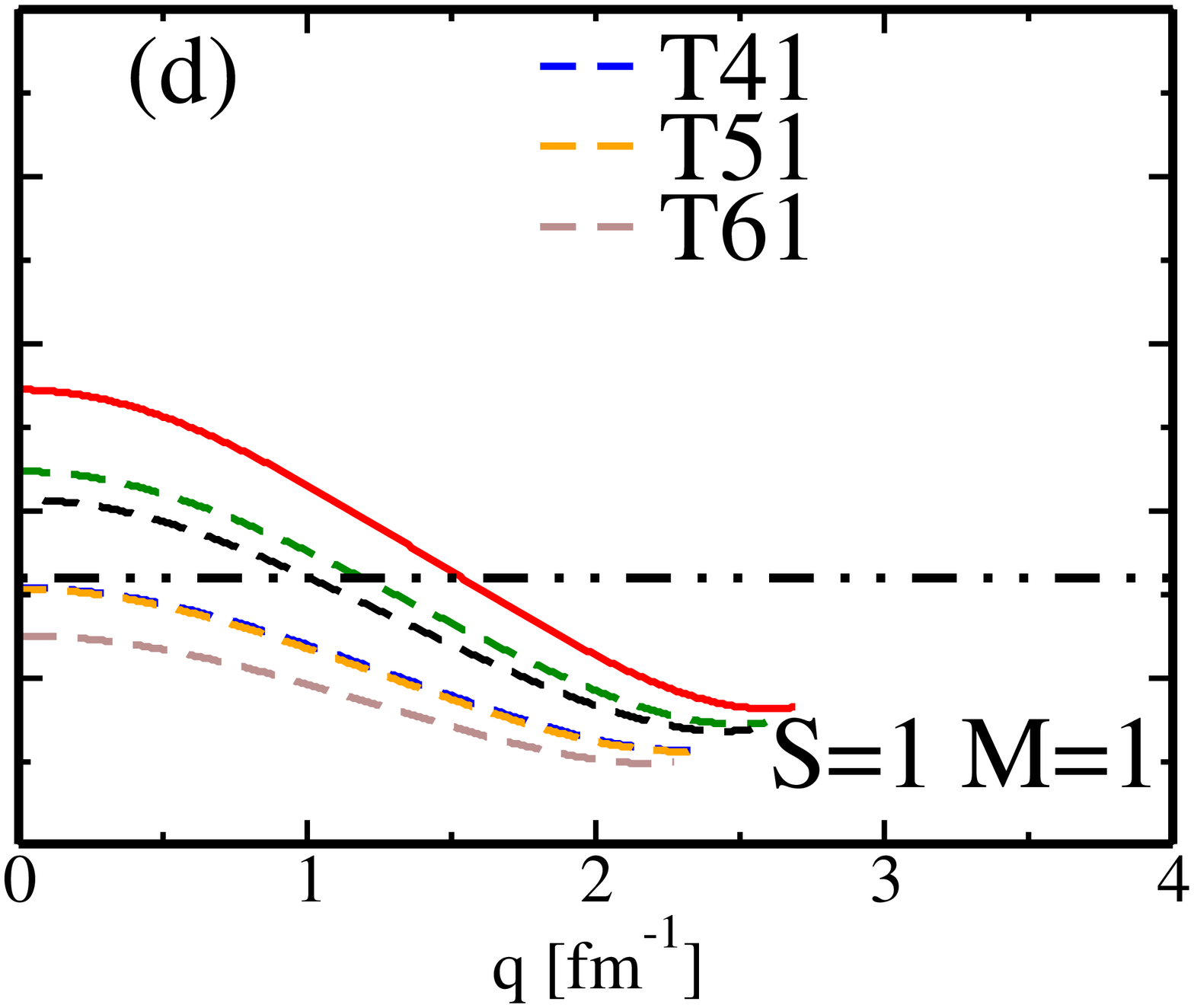} 
\end{matrix}$
\ec
      \caption{(Color on line) For the two $S=1$ channels in PNM, 
               critical densities (in fm$^{-3}$) are plotted
               as functions of the transferred momentum 
               $q$ (in fm$^{-1}$) for the TIJ family of Skyrme forces. 
                The horizontal dashed-dotted line is placed at $\rho=0.16$~fm$^{-3}$ just to guide the eye.}
\label{fig:critical_Tij_S1}
\end{figure}


\begin{figure}[htb]
\bc
$\begin{matrix}
      \includegraphics[clip,scale=0.16,angle=-90,bb=0 0 612 720]{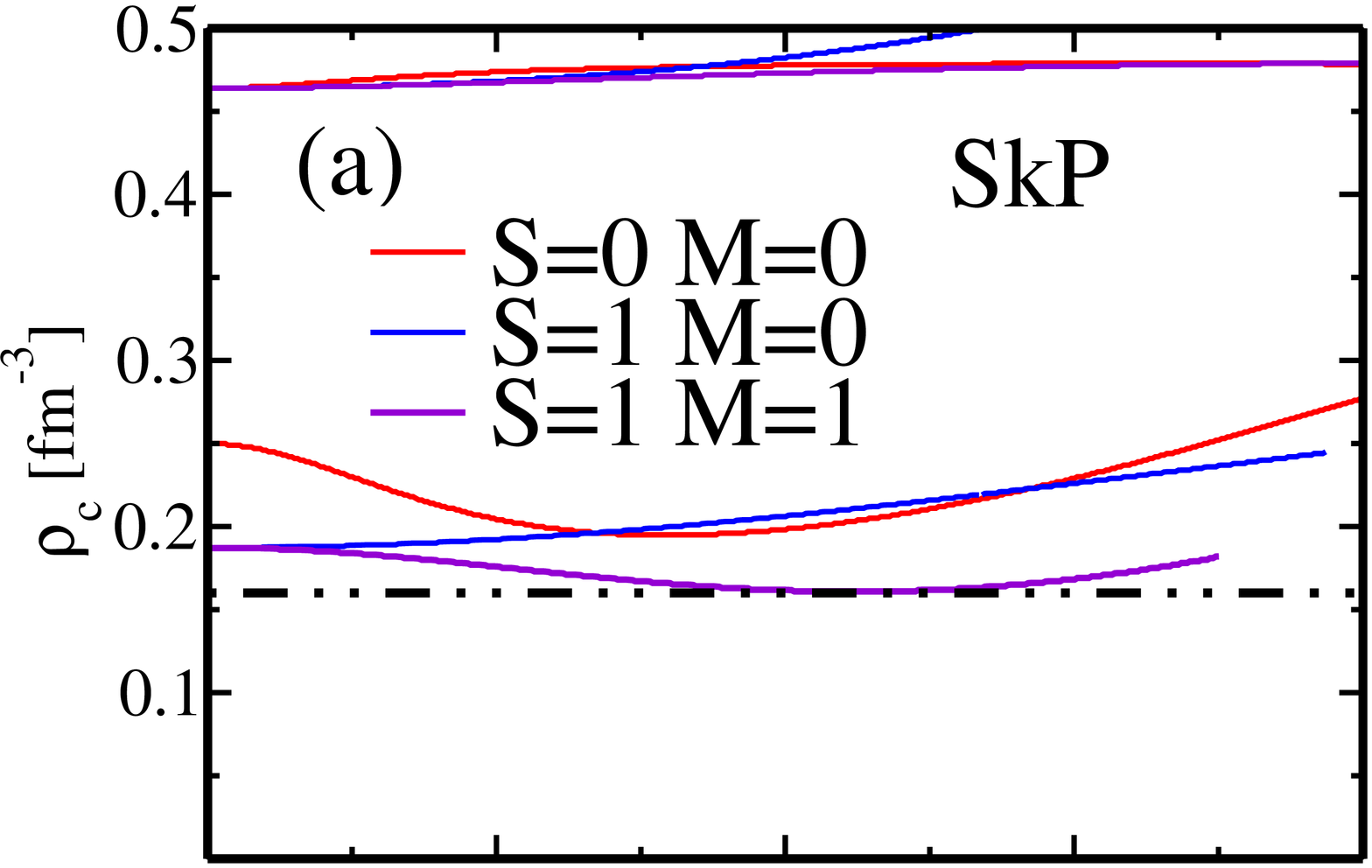} &
      \includegraphics[clip,scale=0.16,angle=-90,bb=0 0 612 720]{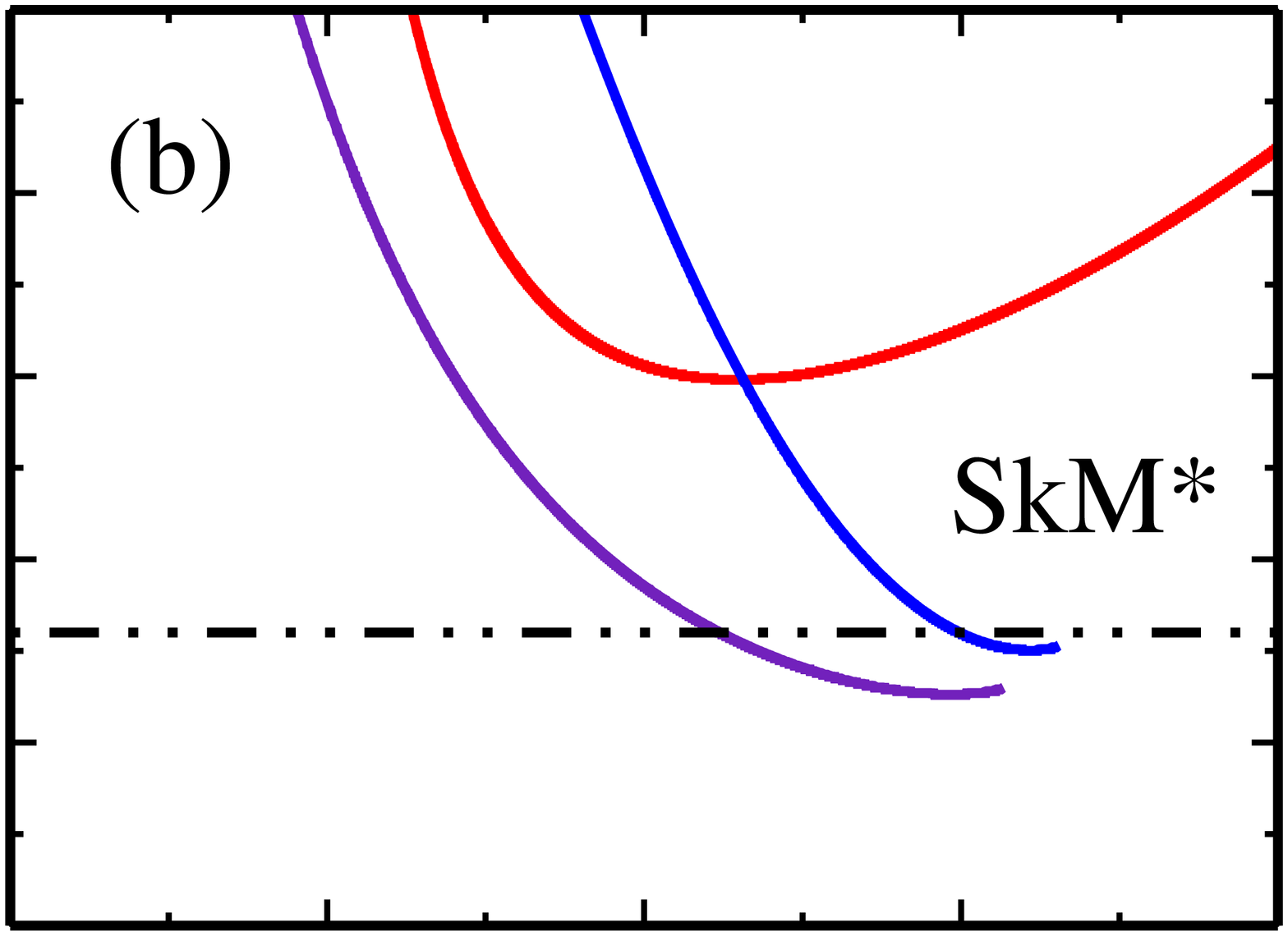} \\
      \includegraphics[clip,scale=0.16,angle=-90,bb=0 0 612 720]{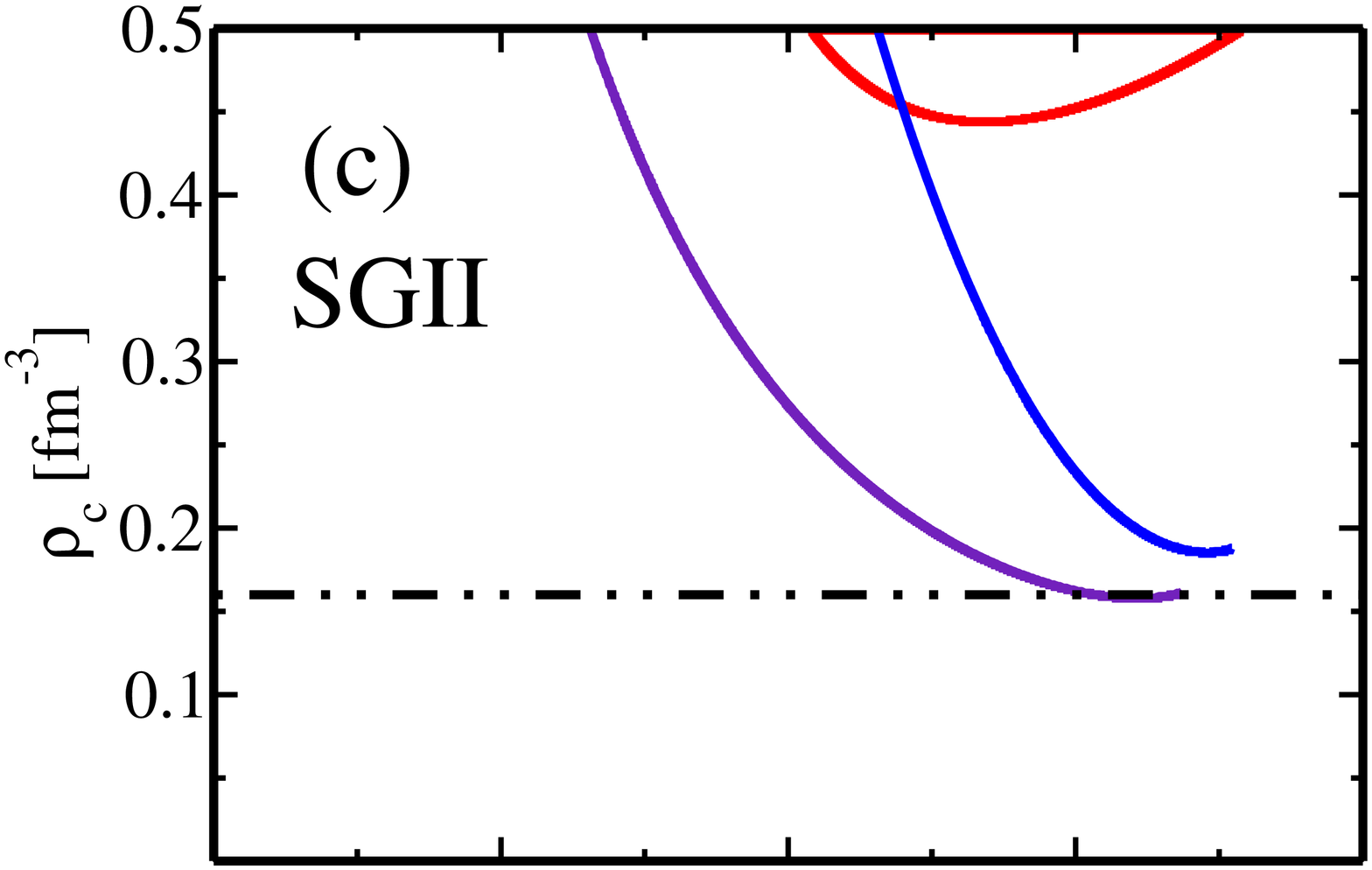} &
      \includegraphics[clip,scale=0.16,angle=-90,bb=0 0 612 720]{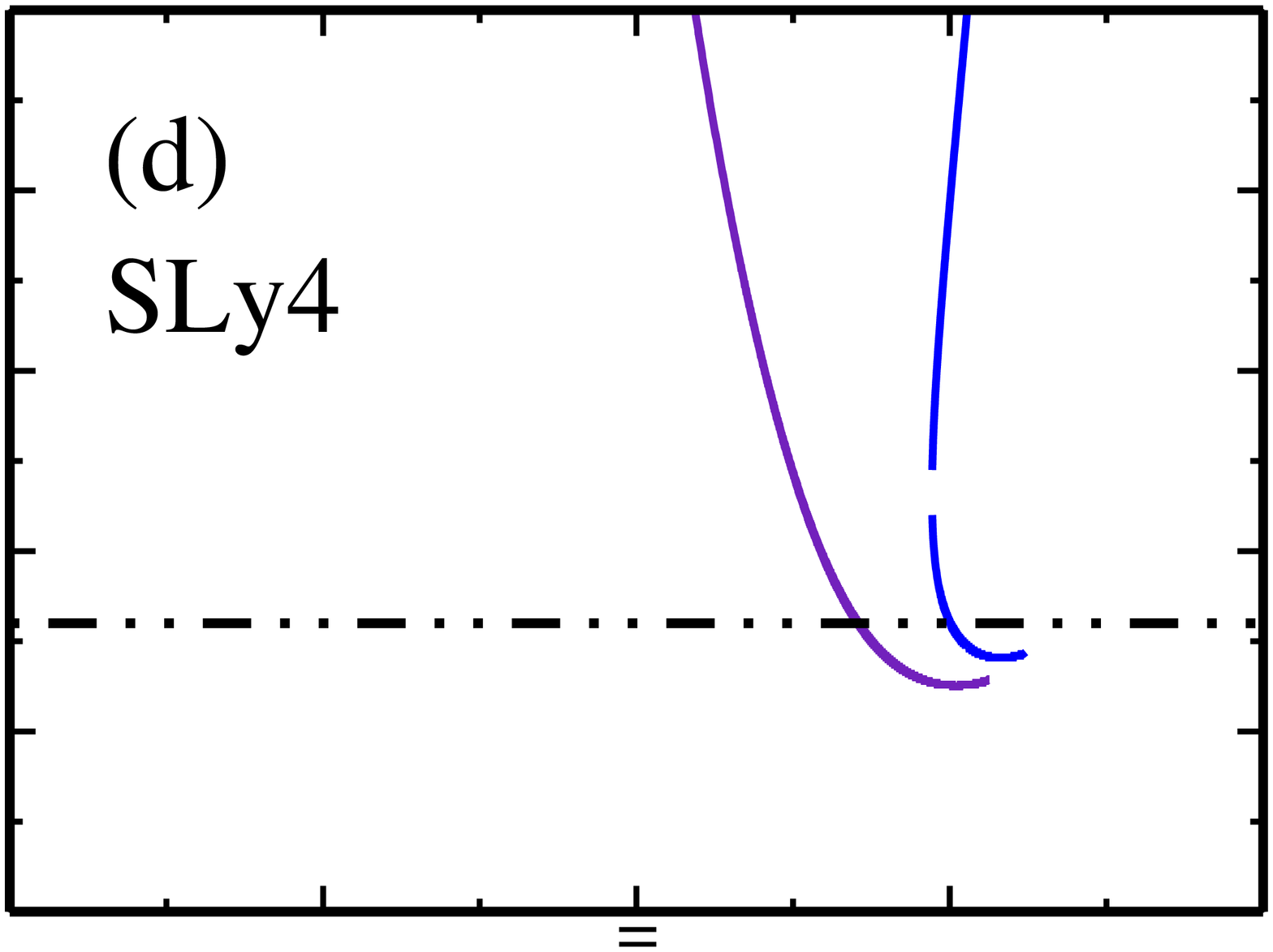} \\
      \includegraphics[clip,scale=0.16,angle=-90,bb=0 0 612 720]{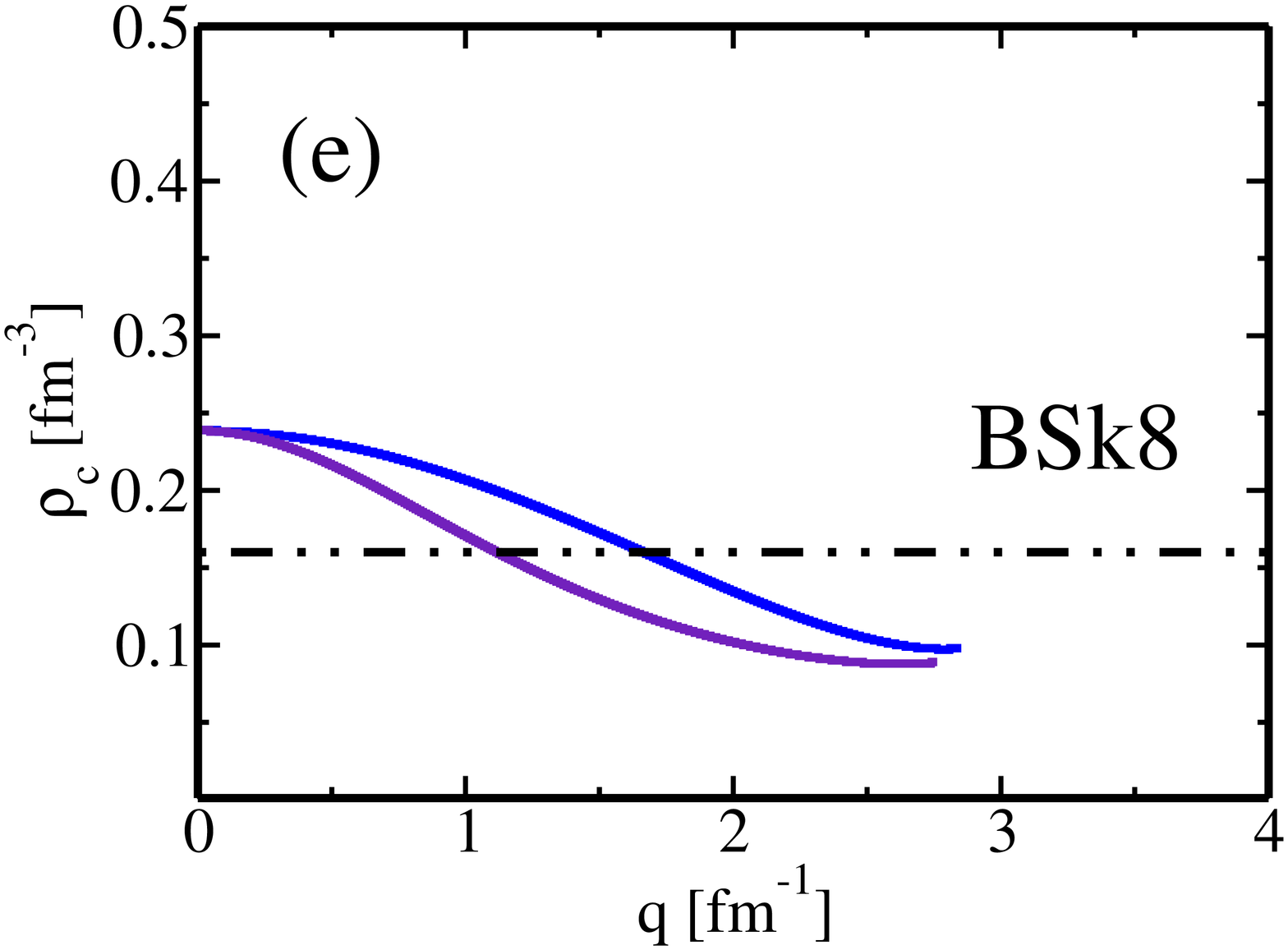} &
      \includegraphics[clip,scale=0.16,angle=-90,bb=0 0 612 720]{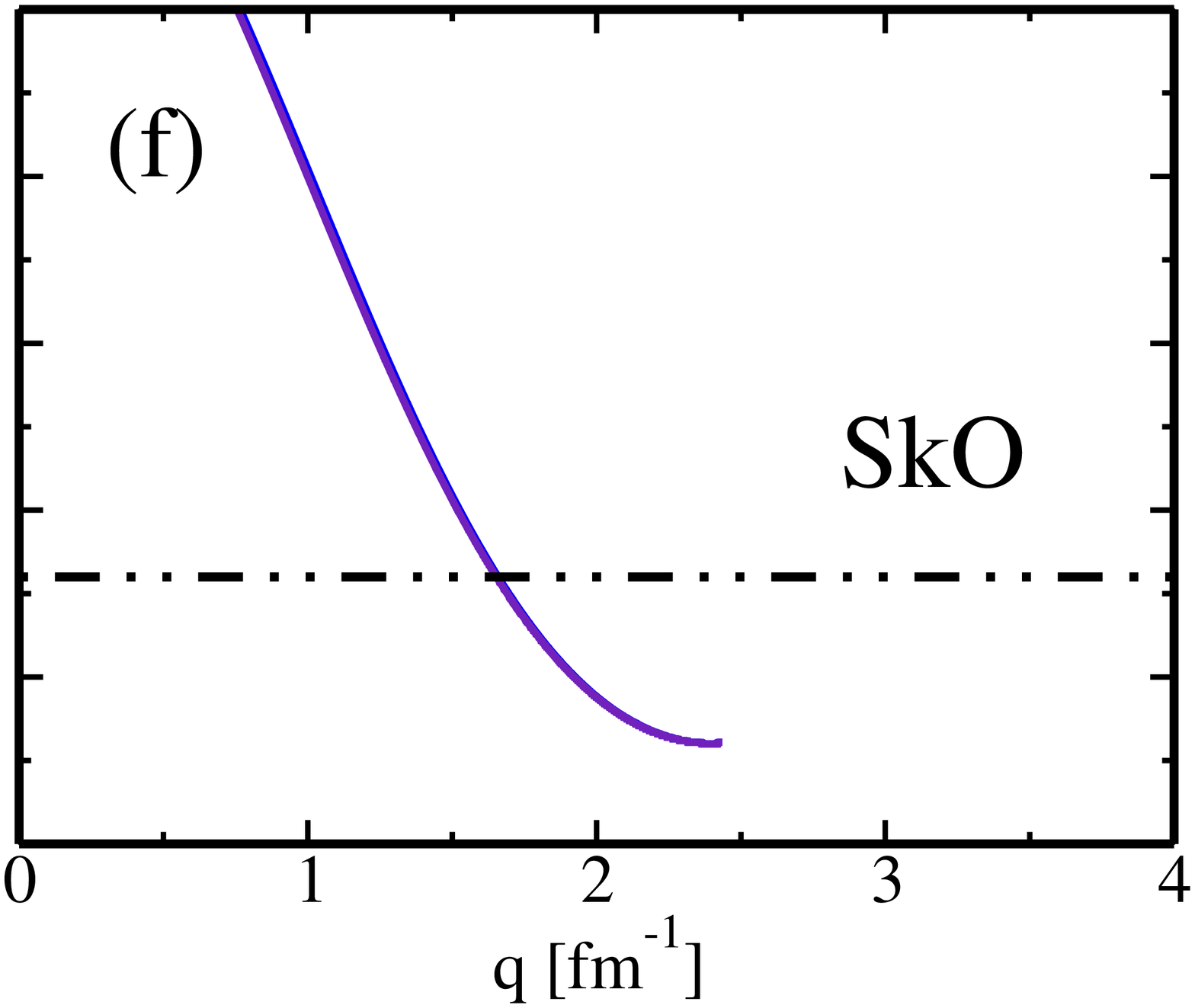} 
\end{matrix}$
\ec
      \caption{(Color on line) For the three PNM channels, 
               critical densities (in fm$^{-3}$) are plotted
               as functions of the transferred momentum 
               $q$ (in fm$^{-1}$) for some usual Skyrme EDF: 
               SkP~\cite{Dob84a}, SkM*~\cite{Bar82a},   SGII~\cite{Giai81}, SLy4~\cite{Cha97a,Cha98a,Cha98b},BSk8~\cite{Samyn05}  and SkO~\cite{Reinhard99}. The horizontal dashed-dotted line is placed at $\rho=0.16$~fm$^{-3}$ just to guide the eye.}
\label{fig:critical_Skyrme}
\end{figure}


Following the notations of article II, we can calculate the most relevant odd-power sum rules for PNM, in particular the $M_{1}$ Energy Weighted Sum Rule (EWSR), the $M_{3}$  Cubic Energy Weighted Sum Rule (CEWSR) and the $M_{-1}$ Inverse Energy Weighted Sum Rule (IEWSR) defined as
\be \label{integral}
M^{(\alp)}_k(q) \, = \, \int_0^{\infty} \mathrm d \omega \, \omega^k \, S^{(\alp)}(q,\omega)\,.
\ee

As stated previously, all the expressions given below are derived for the general Skyrme EDF given
by Eq.(\ref{eq:EF:full}) in which all the coupling constants could be considered as independent ones from the others.
We refer to article II for a detailed discussion on their derivation.

The EWSR in each channel reads
\begin{eqnarray}
M^{(0,0)}_{1}&=&\frac{q^{2}}{2m^{*}}\left(1- W_{2}^{(0)}m^{*}\rho \right)\,,\\
M^{(1,0)}_{1}&=&\frac{q^{2}}{2m^{*}}\left[1- (W_{2}^{(1)}+2C^{F})m^{*}\rho \right]\,,\\
M^{(1,\pm1)}_{1}&=&\frac{q^{2}}{2m^{*}}\left(1- W_{2}^{(1)}m^{*}\rho \right)\,.
\end{eqnarray}

\noindent Taking into account the $W_{2}^{(0)}$ value as well as the neutron effective mass $m^{*}$, defined as
\begin{equation}\label{mass}
\frac{m}{m^{*}}=1+2m(C^{\tau}_{0}+C^{\tau}_{1})\rho,
\end{equation}

\noindent the $M^{(0,0)}_{1}$ EWSR reduces to the free value $q^{2}/2m$, as it should.
As in article II for the case of a Skyrme force, these $M_{1}$ moments can be obtained from the double commutator method~\cite{Bohigas,Lipparini}.

For the CEWSR we have
\bwt
\bqr
\label{eq:m_3}
M^{(0,0)}_{3}   & = & q^4 \frac{k_F^2}{2m^{*3}} \left[ 1 - \mrho \wzd  \right]^2    \nn \\
         & \times & \left\{ \tfrac{3}{5} + k^2 + k^2 \mrho \wzd  
                                         + \frac{\mrho}{2k_F^2} \left[ \wzu + 2 k_F^2 \wzd \right] 
                    \right\} \,, 
                    \eqr
  \bqr
M^{(1,0)}_{3} & = & q^4 \rho^2 [ C^F ]^2 \frac{k_F^2}{5tm^*} 
                      \left\{ 2 \mrho \left[ \wud  + 2 C^F \right] - 1 \right\}      +  q^4 \frac{k_F^2}{2m^{*3}}
                      \left\{ \mrho \left[ \wud  + 2 C^F \right] - 1 \right\}^2                    \nn \\ 
         & \times & \left\{ \tfrac{3}{5} + k^2 + \tfrac{6}{5} m^* \rho_{} C^F 
                                           + k^2 \mrho  \wud  
                                           + \frac{\mrho}{2 k_F^2} 
                                             \left[ \wuu + 4 q^2  C^{\nabla s}
                                                         + 2 k_F^2 \wud \right]      \right\}   \,,  \eqr
  \bqr
M^{(1,\pm 1)}_{3} & = & q^4 \rho_{}^2 \left[ C^F \right]^2 \, \frac{k_F^2}{10 m^*} \, 
                         \left\{ 2 \mrho \wud - 1 \right\}                      + q^4 \frac{k_F^2}{2 m^{*3}} \left[ \mrho  \wud  - 1
                         \right]^2                                         \nn \\
             & \times & \left\{ \tfrac{3}{5} + k^2 + \tfrac{2}{5} \mrho C^F
                                               + k^2 \mrho \wud
                                               + \pdemi \frac{\mrho}{k_F^2}
                                                        \left[ \wuu+ 2 k_F^2 \wud \right]
                        \right\}                                                                       \,.
\eqr

And finally for the IEWSR we have
\bqr\label{eq:m_-1:00}
M^{{\rm (0,0)}}_{-1} & = & f(k) \, \frac{3m^*}{2k_F^2} \, \left\{ 
 - 24 k^2 \left[ \mrho C^{\nabla J} \right]^2 \,
 \frac{f(k) \left[ 1 - 3 \left( k^2 - 1 \right) f(k) \right]}
      {4 - \mrho \left[ 1 - 3 \left( k^2 - 1 \right) f(k) \right]  
                 \left[ \wud  - C^F \right]} \right.                                      \nn \\
                 & - & \tfrac{3}{16} \left[ \mrho f(k) \left( k^2 - 1 \right) \wzd \right]^2    \nn \\
                 & + & f(k) \left[ \frac{k_F m^*}{2\pi^2} \wzu 
                                 + \tfrac{3}{2} \mrho \left( 1 - k^2 \right) \wzd                               
                                 - \tfrac{1}{8} \left( 3 + 13 k^2 \right) \left[ \mrho \wzd \right]^2
                            \right]     +\left. \left[ 1 + \tfrac{3}{4} \mrho \wzd \right]^2 
                                                           \right\}^{-1}             \,,
\eqr
\bqr\label{eq:m_-1:10}
M_{-1}^{(1,0)} & = & f(k)\, \frac{3m^*}{2k_F^2} \, \left\{ 
\left[ 1 + \tfrac{1}{4} \mrho \left( 3 \wud + 4 C^F \right) \right]^2 \right.                            \nn \\
               & - & \tfrac{3}{16} \left[ \mrho f(k) \left( k^2 - 1 \right) \right]^2 \left[ \wud \right]^2   \nn \\
               & + & f(k) \left[ \frac{k_F m^*}{2\pi^2} 
                            \left[ \wuu+ 4 q^2  C^{\nabla s}  \right]  
        - \tfrac{3}{2} \mrho \left( 4 k^2 C^F  + \left( k^2 - 1 \right) \wud \right) \right. \nn \\
               & - & \tfrac{1}{8} m^{*2} \rho^2 \left. \left. \left( 24 ( 1 +    k^2 ) [ C^F ]^2 
                                                                       + 12 ( 1 +  3 k^2 )   C^F \wud
                                                                       +    ( 3 + 13 k^2 ) [ \wud ]^2 \right) 
      \phantom{\frac{1}{1}} \right]                  \right\}^{-1}                                   \,,
\eqr
\bqr\label{eq:m_-1:11}
M_{-1}^{(1,\pm1)} & = & f(k) \, \frac{3m^*}{2k_F^2} \, \left\{
- 12 \left[ \mrho C^{\nabla J} \right]^2 
\frac{ k^2 f(k) \left[ 1 + 3 f(k)( 1 - k^2 ) \right]}
     {4 - \mrho \left[ 1 + 3 f(k)( 1 - k^2 ) \right]  \wzd}              \right.  \nn \\
                  & + & \left[ 1 + \tfrac{1}{4} \mrho \left( 3 \wud + C^F \right) \right]^2 
                    - \tfrac{3}{16} \left[ \mrho f(k) ( 1 - k^2 ) \right]^2 
                      \left( 5 \left[ C^F \right]^2 + 2 C^F \wud + \left[ \wud \right]^2 \right)       \nn \\
                  & + & f(k) \left[ \frac{k_{F} m^*}{2\pi^2} 
                               \wuu 
                                   + \tfrac{3}{2} \mrho ( 1 - k^2 ) \left[ \wud + C^F \right] 
                                                                           \right.  \nn \\
                  & + & \left. \left. \tfrac{1}{8} m^{*2} \rho^2 \left( [ C^F ]^2 ( k^2 - 9 ) 
                                                          - 8 k^2 C^F \wud 
                                                          - ( 3 + 13 k^2 ) [ \wud ]^2 \right)
                               \right]                  \right\}^{-1}                             \,.
\eqr

\ewt
where the $W_{i=1,2}^{(\text{S})}$ coefficients are given in~\citeAppendix{app:rf_pnm}.

As in article II we define the function $f(k)$ as 
\be
f(k) =  \frac{1}{2} \, \left[ 1 + \frac{1}{2k} \left( 1 - k^2 \right) 
                              \ln \left( \frac{k+1}{k-1} \right) \right],
\ee
while $\rho$ and $k$ are now defined for PNM with Fermi momentum $k_{F}$, hence
  
\bqr
    \rho & = & \frac{1}{3\pi^2} k_{F}^3  \,,  \nn \\  
    k      & = & \frac{q}{2k_{F}}          \,.
\nn 
\eqr                              

As already illustrated in the previous section, 
the main effect of the tensor terms is in the $S=1$ channel where one can even observe a divergence at zero energy, but finite transferred momentum.
As explained in detail in the article II, the IEWSR, $M_{-1}$, can be used to detect these poles. 
As an example, on \citeFigure{fig:m-1_T16}, we plotted the IEWSR for T16 obtained from the analytic expansion of the response function (see Eqs.(\ref{eq:m_-1:00})-(\ref{eq:m_-1:11})) 
and from the direct numeric integration. 
We observe on this figure that in the channels $S=1,\,M=0$ and $S=1,\,M=1$ the IEWSR is violated. 
This indicates the presence of a pole in the response function as shown for example for $S^{(1,0)}(q,\omega)$ on figure~\ref{fig:T16} for  $\rho=0.16$~fm$^{-3}$ and transferred momentum $q=1.051$~fm$^{-1}$.



This connection between the pole 
(when it does exist) observed in the response function and the pole observed
in the $M_{-1}$ sum rule has been discussed in article II and we refer to it for a more detailed discussion on this point.
It is thus possible to determine in a systematic way the critical density at which a pole occurs for a given momentum $q$ from Eqs.(\ref{eq:m_-1:00})-(\ref{eq:m_-1:11}).
On \citeFigure{fig:critical_T16} we display such critical 
densities, $\rho_{c}$ with respect to the transferred momentum for the T16 interaction.
On the left panel we first show the position of the poles of the response function for SNM for each ${\rm (S,M,I)}$ channel. On the right panel we then show the position of the poles for PNM for each ${\rm (S,M)}$ channel.

Even if we exclude the case of spinodal instability which is not present in PNM, one can see that the presence and the location of the poles depends strongly on the system under analysis: for a given interaction, the critical densities are very different for PNM and SNM. Similarly we show in Fig. \ref{fig:critical_Tij_S1} the critical densities for the tensor parameterizations that we use in the following section to study neutrino mean free path.

Following  article II, we show for completeness in \citeFigure{fig:critical_Skyrme} the critical densities for the Skyrme EDF previously analyzed in article II for SNM, but in this case for PNM.
We observe that SkP behaves very differently from the other Skyrme forces presented in this article. It presents a first  instability in the $S=0$ channel at   $\rho\approx0.16$ fm$^{-3}$ and a second one at higher density $\rho\approx0.45$ fm$^{-3}$ due to the presence of a pole in the effective mass defined
in Eq.~(\ref{mass}).

\section{Neutrino mean free path}
\label{sect:lbd_neutrino}

The aim of this section is to investigate the effect of the choice of the parameters
of the tensor terms on the neutrino mean free path in neutron matter.
In this article we restrict ourselves to the academic case of neutron matter at zero temperature (the generalisation at finite temperature $T$ is in progress). 
This, of course, implies some restriction on the application of our approach to neutron stars studies. 
At first stages of the cooling process of the neutron stars, modelized as asymmetric nuclear matter, the high temperatures involved allow charged current reactions; then, 
when the temperature decreases, neutral currents dominate. 
Since we consider the case $T=0$ and the pure neutron matter, we will take into account only neutral currents for the determination of the neutrino mean free path. This quantity is defined as
\begin{equation}
\lambda=(\sigma \rho)^{-1},
\end{equation}
where $\sigma$ is 
the total cross section for the neutral current reaction 
$\nu + n \longrightarrow \nu' + n'$. In the absence of tensor (and spin-orbit) interaction this 
total cross section is obtained by integrating the double differential cross-section 
per neutron 

\begin{align}
\frac{d^2 \sigma (E_\nu)}{d \Omega_{k'} d \omega} = 
\frac{G_F^2 E_\nu'^2}{16\pi^2}& \left[(1+\cos \theta)R^V(q,\omega)\right.\nonumber \nonumber\\
&\left.+g_A^2(3-\cos \theta)R^A(q,\omega)\right]\,,
\label{eq_d2s_notens}
\end{align}
where $ G_F $ is the weak coupling constant, $g_A=1.255$~\cite{Martini:2009uj} is the axial charge of the nucleon,  
$E_\nu$ and $E_\nu'$ are the incoming and outgoing neutrino energies, $\theta$ the scattering 
angle between the incoming and outgoing neutrino momenta and $\omega = E_\nu- E_\nu'$ and 
$\vec{q}= \vec{k}-\vec{k'}$ the energy and the momentum transfer in the reaction. The response functions $R^V(q,\omega)$ and $R^A(q,\omega)$ describe the response of the system to density fluctuations ($S=0$) and spin fluctuations ($S=1$) related to the coupling of neutrino 
to vector and axial currents of the neutron. These response functions are defined as
\begin{eqnarray}
R^V(q,\omega)&=&-\frac{1}{\pi \rho} \textrm{Im}\chi^{(S=0)}(q,\omega)\,,\nonumber \\
R^A(q,\omega)&=&-\frac{1}{\pi \rho} \textrm{Im}\chi^{(S=1)}(q,\omega)\,.
\end{eqnarray}
When tensor forces are considered the spin response is splitted into two components, {\em i.e.} the spin
longitudinal response
\begin{equation}
R^A_L(q,\omega)=-\frac{1}{\pi \rho} \textrm{Im}\chi^{(S=1,M=0)}(q,\omega)
\end{equation}
and the spin transverse response
\begin{equation}
R^A_T(q,\omega)=-\frac{1}{\pi \rho} \textrm{Im}\chi^{(S=1,M=\pm1)}(q,\omega).
\end{equation}
These responses can be considerably different one of the other and compared to
the case without tensor interaction.
This has important consequences on the neutrino cross sections and
the neutrino mean free path. 

\begin{figure}[ht]
\begin{center}
  \includegraphics[clip,scale=0.339]{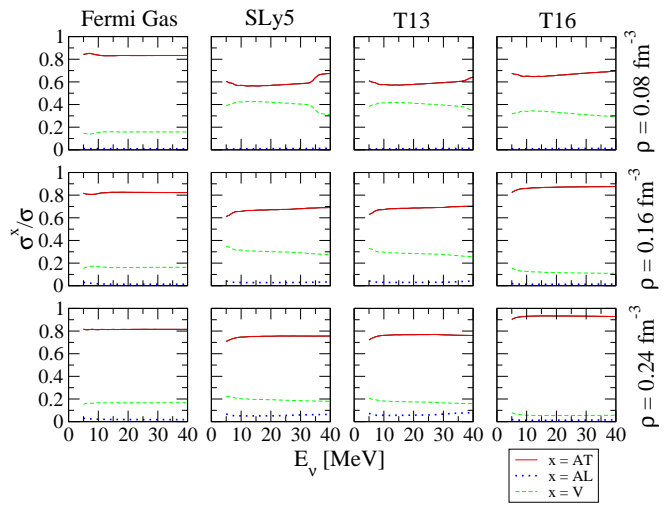}
\caption{(Color online) Axial spin transverse (AT), axial spin longitudinal (AL) and vector (V) 
contributions to the cross section per neutron for the reaction $\nu + n \longrightarrow \nu' + n'$ in the neutron matter. Three different densities as well three different interactions are considered. The Fermi gas result is also shown.}
\label{fig_ratio_sigma}
\end{center}
\end{figure}
In the presence of tensor interaction the double differential cross-section 
per neutron for neutral current reaction is given by
\begin{align}
&\frac{d^2 \sigma (E_\nu)}{d \Omega_{k'} d \omega} = 
\frac{G_F^2 E_\nu'^2}{16\pi^2} \left\{\phantom{\frac{1}{1}}(1+\cos \theta)R^V\right.\nonumber\\
&+g_A^2\left[\frac{2 (E_\nu'\cos \theta-E_\nu)(E_\nu'-E_\nu \cos \theta)}{q^2}+1 -\cos \theta\right]
R^A_L\nonumber\\
&+\left.g_A^2~2 \left[\frac{E_\nu E_\nu'}{q^2} \sin^2\theta+1-\cos \theta \right]R^A_T\right\}\,.
\label{eq_d2s_tens}
\end{align}
This expression reduces to Eq.~(\ref{eq_d2s_notens}) when $R^A_L=R^A_T=R^A$ as one can easily observe remembering that for neutral current
\begin{eqnarray}
q^2&=&(\vec{k}-\vec{k'})^2\nonumber\\
&=&k^2+k'^2-2\vec{k}\cdot\vec{k'}\nonumber\\
&=&E_\nu^2+ E_\nu'^2-2E_\nu E_\nu' \cos \theta\,.
\end{eqnarray}

\begin{figure}[t]
\bc
$\begin{matrix}
      \includegraphics[clip,scale=0.16,angle=-90]{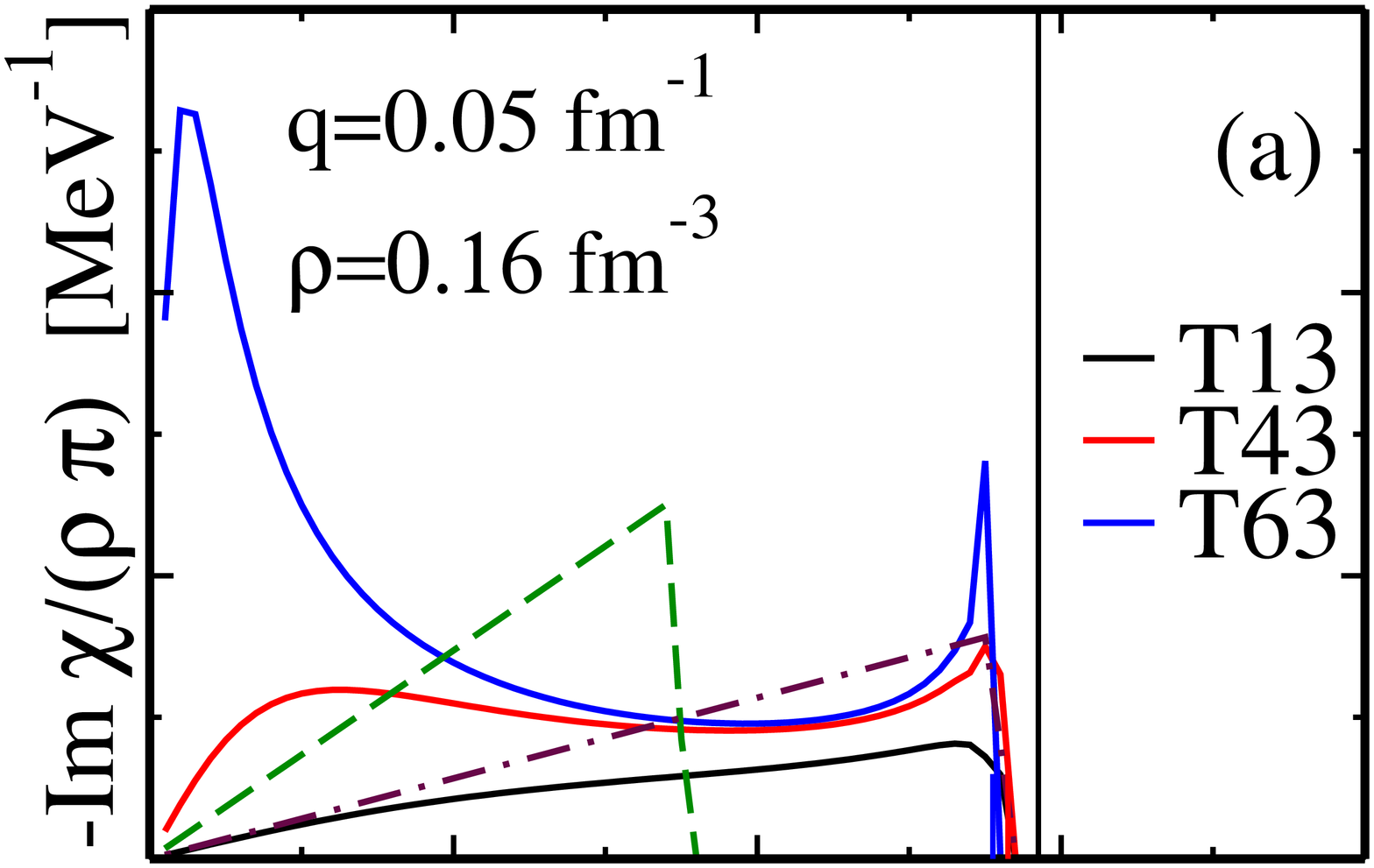} &
      \includegraphics[clip,scale=0.16,angle=-90]{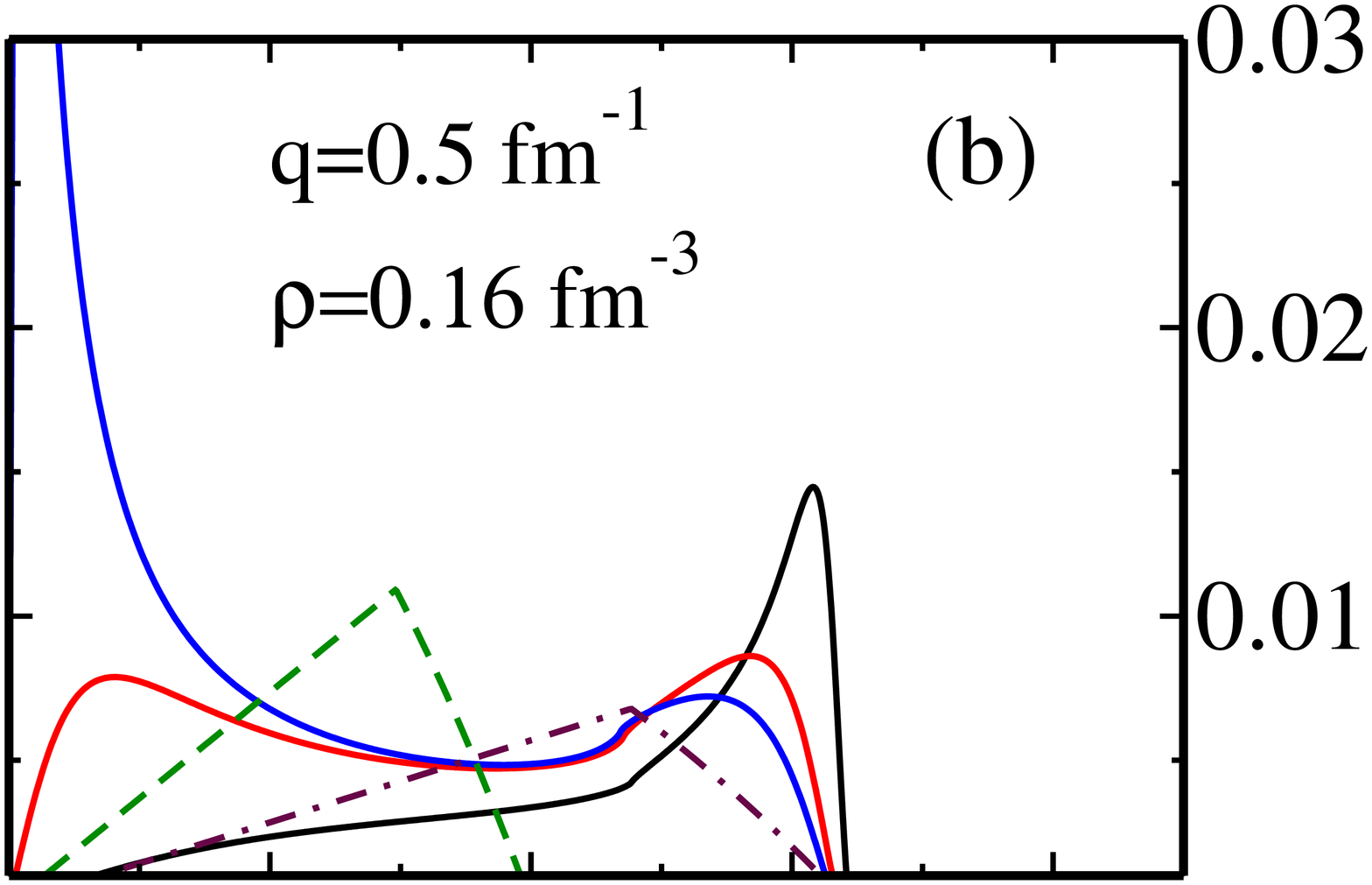} \\
      \includegraphics[clip,scale=0.16,angle=-90]{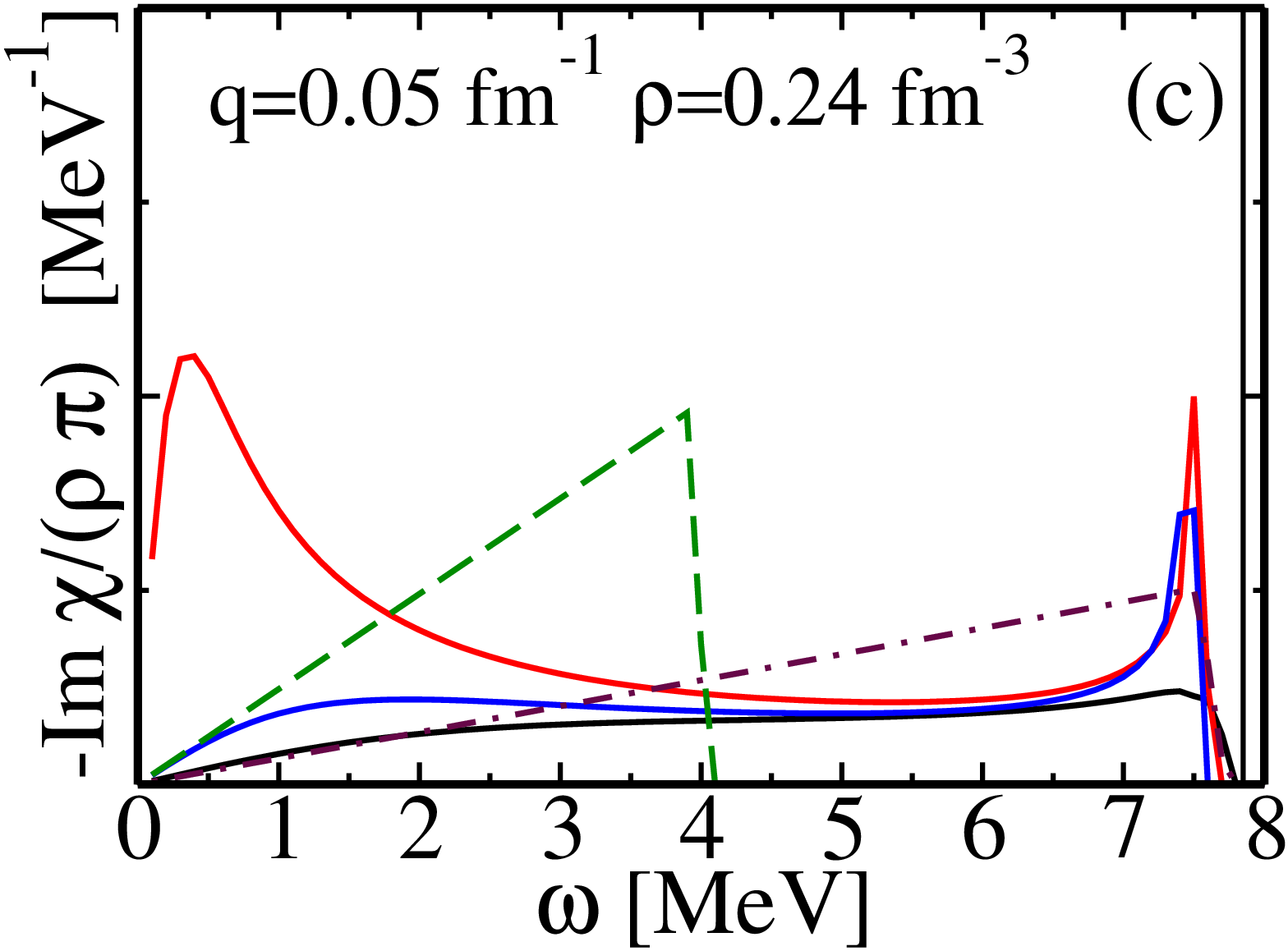} &
      \includegraphics[clip,scale=0.16,angle=-90]{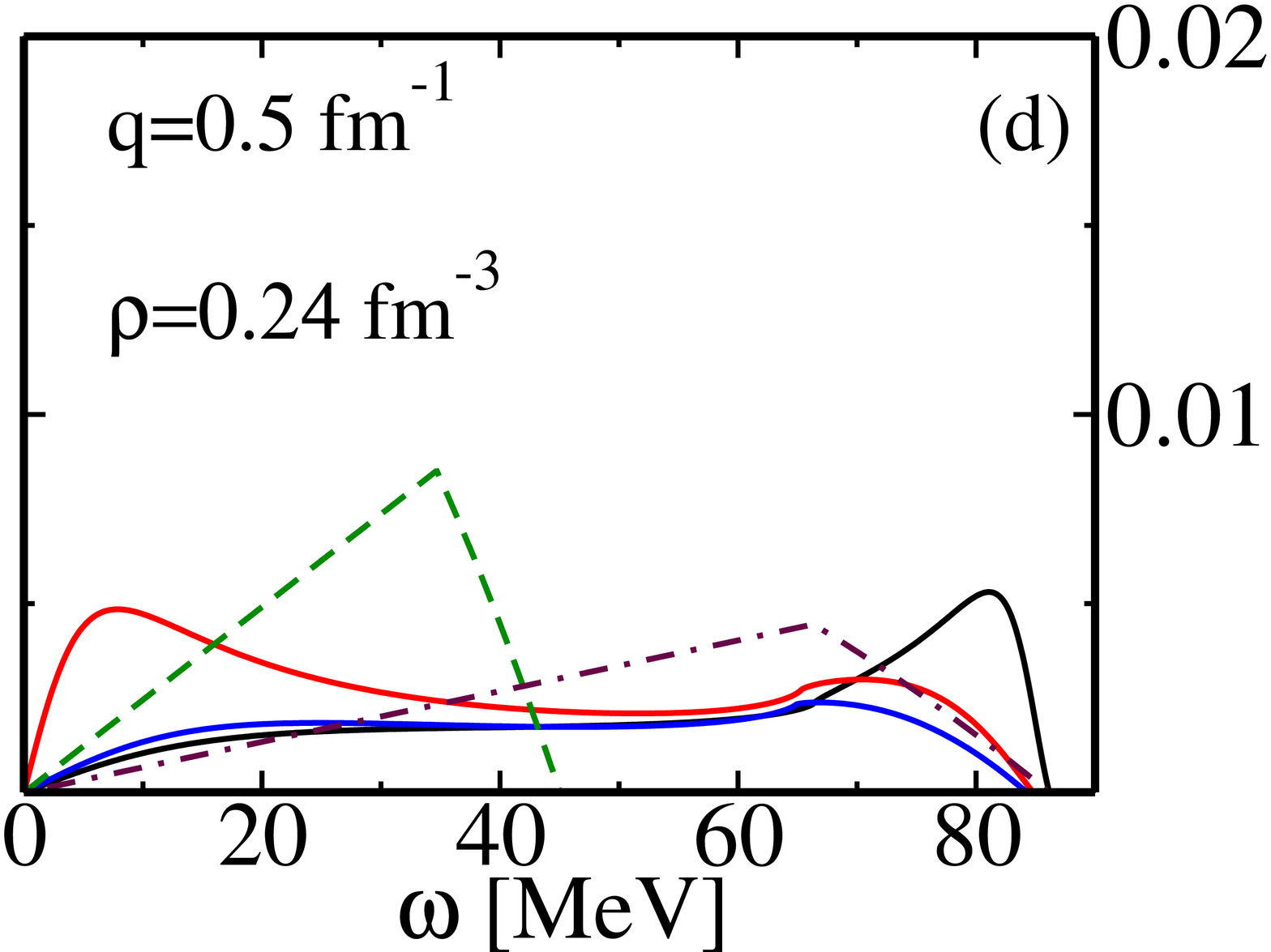} 
\end{matrix}$
\ec
      \caption{(Color on line) The response functions
 for the channel $S=1,M=1$ for three different forces (T13, T43 and T63) is shown with solid lines. On the same figure we
represent the Fermi gas (dashed line) and the uncorrelated response functions (dashed dotted), {\em i.e.} when the residual
interaction is put to zero, for the interaction T13. The uncorrelated
response functions are not equivalents among T13, T43 and T63 due to
the small differences in the effective mass. }
\label{fig:marco2}
\end{figure}

In Eq. (\ref{eq_d2s_tens}), as well as in Eq. (\ref{eq_d2s_notens}), we neglect corrections of order $\frac{E_\nu}{m}$ from weak magnetism and 
other effects~\cite{Horowitz:2001xf} like the finite size of the nucleon or nucleon excitations~\cite{Chen:2009am}. 
A generalization of Eq. (\ref{eq_d2s_tens}) taking into account of all these effects can be found for example in Ref.~\cite{Martini:2009uj}. 
As already stressed in Ref.~\cite{Martini:2009uj} (and illustrated in Ref.~\cite{Martini:2010ex} for charged current reaction) the cross section is dominated by 
the spin transverse response $R^A_T$. In Fig.~\ref{fig_ratio_sigma} we present the relative axial spin transverse, axial spin longitudinal and vector contributions to the 
neutral current cross section. We consider four different cases. The first is the Fermi gas.
In this case $R^A_L=R^A_T=R^A$ so the difference between 
the three contributions is only due to the  kinematical factors and the coupling constants multiplying the response functions. 
Second we consider the case when the response functions are calculated with the SLy5 force.
In this case the coupling constants $C^{F}=C^{\nabla s}=0$, purely related with the tensor part of
the interaction, do not contribute. A 
possible difference between  $R^A_L$ and $R^A_T$ is due to the spin orbit contribution and according to the Eqs.\ref{bbw00} and \ref{bbw11}
given in Appendix \ref{app:rf_pnm}, the $q^4$ factor 
reduces these differences at low momentum transfer. Finally we consider the T13 and T16 parameterizations of tensor contributions.
The results for three different densities, $\rho= 0.08$, 0.16 and 0.24~fm$^{-3}$ are also shown on
Fig.~\ref{fig_ratio_sigma}.
As it clearly appears all the cross sections are dominated by the spin transverse response but this contribution may vary between $\sim$\,60\,\% and $\sim$\,90\,\% 
depending on the interactions and densities considered. 
It reflects the possible quenching, enhancement or divergence of the nuclear response functions. 
This behavior was already illustrated in the section \ref{subsec_lr} in connection with the Fig. \ref{fig:marco1}. 
To complete our discussion we plot in Fig. \ref{fig:marco2} the spin transverse response for $q=0.05$~fm$^{-1}$ and 
$q=0.5$~fm$^{-1}$, for $\rho=0.16$~fm$^{-3}$ and $\rho=0.24$~fm$^{-3}$ and for the tensor forces T13, T43 and T63.
A collective spin zero sound characterizes the $S=1,\,M=\pm1$ 
response for T13 at $q=0.05$~fm$^{-1}$ for $\rho=0.16$~fm$^{-3}$ and $\rho=0.24$~fm$^{-3}$. 
A similar behaviour, with the corresponding quenching of the p-h continuum characterizes the 
$S=1,\,M=\pm1$ responses for SLy5 and and T16 as shown on
Fig.~\ref{fig:marco1} 
for $\rho=0.08$~fm$^{-3}$ and $\rho=0.16$~fm$^{-3}$. The T63 force on the other hand is characterized by 
an enhancement of the response at low $\omega$ for $\rho=0.16$~fm$^{-3}$. 
At $q=0.5$~fm$^{-1}$ this enhancement seems to be critical. At $\rho=0.24$~fm$^{-3}$ 
the enhancement of the T16 response no longer holds. In this case this response is suppressed with respect to the corresponding Hartree-Fock response. An enhancement with respect to the HF and FG case for small $\omega$ characterizes for this density the T43 force. 
\begin{figure}[ht]
\begin{center}
  \includegraphics[clip,scale=0.35]{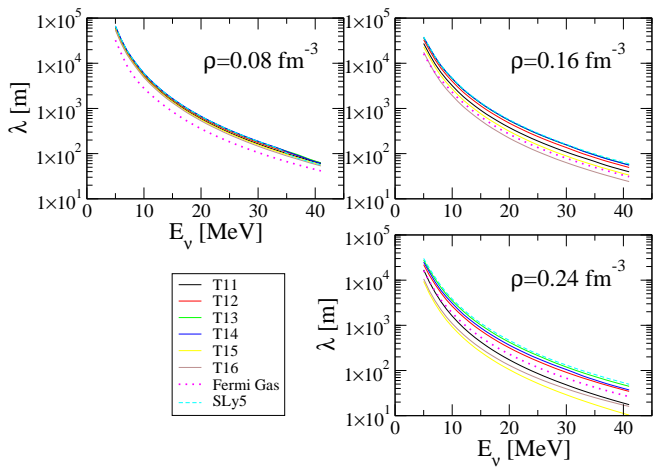}
\caption{(Color online) Neutrino mean free path for scattering reaction $\nu + n \longrightarrow \nu' + n'$ in neutron matter. 
The dotted lines correspond to the non interacting Fermi gas case. The dashed lines are the interacting case with SLy5 force. The continuous lines correspond to 
several parameterizations of the tensor contribution following the T11-T16 chain. The system densities considered in the three panels are $\rho=0.08$, 0.16 and 0.24~fm$^{-3}$.}
\label{fig_lambda_T11_T16}
\end{center}
\end{figure}

\begin{figure}[ht]
\begin{center}
  \includegraphics[clip,scale=0.35]{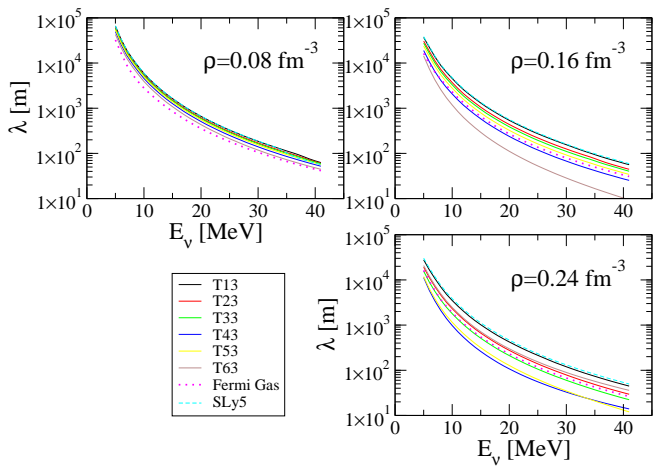}
\caption{(Color online) The same as Fig. \ref{fig_lambda_T11_T16} but for T13-T63 chain.}
\label{fig_lambda_T13_T63}
\end{center}
\end{figure}

These different behaviors affect obviously the neutrino mean free path. 
We have calculated it for T11-T16 and T13-T63 chains of parameterizations of Skyrme 
tensor interactions for a neutrino energy 5 MeV $< E_\nu <$ 40 MeV and for three values 
of densities, {\em i.e.} $\rho=0.08$, 0.16 and 0.24~fm$^{-3}$. 
Note that when a spin zero-sound collective mode appears one must in principle include it in the calculation of the neutrino mean free path. For this mode the response function reduces to a delta distribution. 
Nevertheless, as already observed in~\cite{Iwamoto:1982zp}, the collective mode itself gives little scattering, its contribution is negligible when calculated in the Landau approximation. 
It was also observed in~\cite{Navarro:1999jn} that for all the Skyrme forces there considered, this magnon rapidly disappears with the temperature because of a strong Landau damping. 
Hence we do not explicitly compute the spin zero sound contribution. Its effect is on the other hand present as a suppression of the corresponding p-h continuum and has a consequence on the cross section.

\begin{figure}[ht]
\begin{center}
  \includegraphics[clip,scale=0.3]{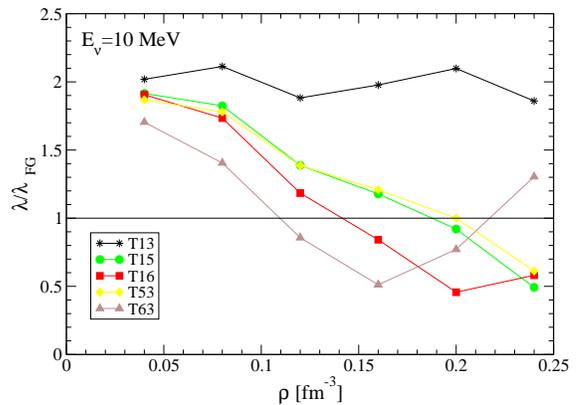}
\caption{(Color online) The relative neutrino mean free path in neutron matter $\lambda/\lambda_{\text{FG}}$ as a function of the density 
for some parameterizations of the tensor terms. The incoming neutrino energy is $E_\nu$=10 MeV.}
\label{fig_lambda_vs_rho_ratio}
\end{center}
\end{figure}

The results for the neutrino mean free path are reported in Fig. \ref{fig_lambda_T11_T16} for the T11-T16
chain and In Fig.~\ref{fig_lambda_T13_T63} for the T13-T63 chain.  
In each figure we include the Fermi gas result and the calculation with the SLy5 force which does not have tensor terms. 
From Figs.~\ref{fig_lambda_T11_T16} and \ref{fig_lambda_T13_T63} clearly emerges the 
crucial dependence of the neutrino mean free path on the parameterizations of 
the Skyrme tensor terms. For $\rho=0.08$~fm$^{-3}$ the different parameterizations give quite similar results 
and higher with respect to the non interacting case. Already at $\rho=0.16$~fm$^{-3}$
the spread becomes important. For some cases $\lambda$ is higher than the Fermi gas one, for other is lower. 
In many cases it is lower than the corresponding SLy5 result. 
Also the behavior of $\lambda$ with the density is not so trivial, as already appears in the three panels of
Figs.~\ref{fig_lambda_T11_T16} and~\ref{fig_lambda_T13_T63}. This is
clearly shown on Fig.\ref{fig_lambda_vs_rho_ratio} where the ratio $\lambda/\lambda_{\text{FG}}$ is plotted.
For example for T13 this quantity stays quite constant 
with the density, for T15 and T53 it decreases and for T16 and T63 it is no longer monotone.

\section{Summary and conclusions} 
In this article we have calculated the RPA response function for pure neutron matter considering Skyrme
energy density functionals including tensor terms. 
This article parallels with~\cite{Davesne09,Davesne12} where similar calculations were
performed for the symmetric nuclear matter. 
As in previous articles divergences and instabilities of the response~\cite{Davesne12,Pastore12b},
in particular in the $S=1$ channel, are discussed in connection with the sum rules. 

We applied our results to the study of the neutrino mean free path. The advantage of the present framework
is that it allows to describe 
nuclear (and neutron) matter equation of state and the neutrino mean free path simultaneously, hence in a
self-consistent way. Obviously, before to achieve reliable description 
of neutrino transport phenomena in neutron stars, the calculations performed here must be generalized
at finite temperature and for asymmetric nuclear matter. Nonetheless, already at this
oversimplified level (pure neutron matter at zero temperature) we have shown the strong dependence of the
neutrino mean free path on the tensor term parameterizations. 
It represents an important reason, among others, to an accurate treatment of Skyrme functionals including
tensor contribution.


\section*{Acknowledgments}

This work was supported by the NESQ project (ANR-BLANC 0407).
The authors thanks  M. Ericson and J. Navarro for  stimulating and encouraging discussions. 
The discussions with T. Duguet, M. Bender and J. Margueron are also acknowledged.
M.M. acknowledges the Communaut\'e Fran\c caise de Belgique (Actions de Recherche Concert\'ees) for financial support.

\newpage

\begin{appendix}
%
%
\section{Particle-hole matrix elements in presence of a zero range tensor 
         interaction.}
\label{app:phme_tensor}

Following the notation adopted in article I and II, we give in Table~\ref{arrayfinale} the values of the particle-hole residual interaction for the tensor part of the functional in terms of the $B^{x}$,  with $x=\Delta s,F,\dots$,  coefficients of the functional. This particular notation has been already discussed in article II and we refer to it for detailed explanations.

\bwt

\begin{table}[h!]
\caption{Contribution of the EDF tensor part to the residual interaction
 in terms of the $B^{x}$ coupling constants. 
For the sake of simplicity we have introduced 
$\bbK_{\rm i,j}=[(k_{12})_{i}(k_{12})_{j}]$, 
where $(k_{12})_{M}^{(1)}$ is defined in Eq.(9) of article I. 
The term $\delta_{SS'}\delta_{S1}$ is implicit everywhere.
\label{arrayfinale}}
\begin{ruledtabular}
\begin{tabular}{cccc} 
                     &&&                                                               \\
                     & $M'=1$ & $M'=0$ & $M'=-1$                                       \\ 
                     &        &        &                                               \\ 
\hline
                     &        &        &                                               \\
\multirow{3}{12mm}{$M=1$} & $-2 q^2 \, \left( B^T + 4 B^{\Delta s} \right)$                               
                          &&                                                           \\ 
                          & $+4 \, B^T \, \bbKzz$               
                          & $-4 \, B^F \, \bbKmz$  
                          & $-4 \, B^F \, \bbKmm$                                      \\ 
                          & $-4 \, \left( 2 B^T + B^F \right) \, \bbKum$              
                          &&                                                           \\
                          &&&                                                          \\ 
\hline
                          &&&                                                          \\ 
\multirow{3}{14mm}{$M=0$} && $-2 \, q^2 \, \left( B^T - 4 B^{\nabla s} + 4 B^{\Delta s} + B^F \right)$
                          &                                                            \\     
                          & $ 4 \, B^F \, \bbKzu$    
                          & $+4 \, \left( B^T + B^F \right) \, \bbKzz$   
                          & $ 4 \, B^F \, \bbKmz$                                      \\
                          &&$-8 \, B^T \, \bbKum$   
                          &                                                            \\
                          &&&                                                          \\ 
\hline
                          &&&                                                          \\   
\multirow{3}{14mm}{$M=-1$}&&& $-2 \, q^2 \, \left( B^T + 4 B^{\Delta s} \right)$       \\ 
                          & $-4 \, B^F \, \bbKuu$         
                          & $-4 \, B^F \, \bbKuz$            
                          & $+4 \, B^T \, \bbKzz$                                      \\
                          &&& $-4 \, \left( 2 B^T + B^F \right) \, \bbKum$             \\
                          &&&                                                          \\

\end{tabular}
\end{ruledtabular}
\end{table}

%
%
\section{Response functions}
\label{app:rf_pnm}

This appendix contains the explicit expressions for the response functions for pure neutron matter. Since the isospin is no longer a relevant quantum number, each channel is denoted as $(1,M)$ for $S=1$ or only $(0)$ for $S=0$.

We have
\bi
   \item for the $S=0$ channel
\bqr
\frac{\chi_{HF}}{\chi_{RPA}^{(0,0)}}&=&1-\widehat{W}_{1}^{(0,0)}\chi_{0}+W_{2}^{(0)}\left( \frac{q^{2}}{2} \chi_{0} -2k_{F}^{2}\chi_{2}\right)\nonumber\\
&+&[{W}_{2}^{(0)}]^{2}k_{F}^{4}\left[ \chi_{2}^{2}-\chi_{0}\chi_{4}+\left( \frac{m^{*}\omega}{k_{F}^{2}}\right)^{2}\chi_{0}^{2}-\frac{m^{*}}{6\pi^{2}k_{F}}q^{2}\chi_{0}\right]+2\chi_{0}\left( \frac{m^{*}\omega}{q}\right)^{2} \frac{{W}_{2}^{(0)}}{1-\frac{m^{*}k_{F}^{3}}{3\pi^{2}}{W}_{2}^{(0)}}\,,
\eqr

   \item for the $S=1$ channels
\bqr
\frac{\chi_{HF}}{\chi_{RPA}^{(1,0)}}&=&\left(1+\frac{k_{F}^{3} m^{*}{C}^{F}}{3\pi^{2}} \right)^{2}+\widehat{W}_{1}^{(1,0)}\chi_{0} \nonumber\\
&+&{W}_{2}^{(1)}\left[\frac{q^{2}}{2}\left( 1+\frac{2{C}^{F}k_{F}^{3}m^{*}}{3\pi^{2}}\right)\chi_{0}-2k_{F}^{2}\chi_{2}+\frac{2k_{F}^{5}m^{*}{C}^{F}}{3\pi^{2}}\left( \chi_{0}-\chi_{2}\right) \right]\nonumber\\
&+&[\breve{W}_{2}^{(1)}]^{2}\left[k_{F}^{4}\chi_{2}^{2}-k_{F}^{4}\chi_{0}\chi_{4}+m^{*2}\omega^{2}\chi_{0}^{2}-\frac{k_{F}^{3}m^{*}q^{2}}{6\pi^{2}}\chi_{0} \right]\nonumber\\
&+&\frac{2m^{*2}\omega^{2}}{q^{2}}\frac{({W}_{2}^{(1)}+2{C}^{F})\left[ 1+\frac{k_{F}^{3}m^{*}}{3\pi^{2}}{X}^{(1,0)}\right]}{1+\frac{k_{F}^{3}m^{*}}{3\pi^{2}}({X}^{(1,0)}-{W}_{2}^{(1)}-2{C}^{F})}\chi_{0}
\eqr
and
\bqr
\frac{\chi_{HF}}{\chi_{RPA}^{(1,\pm1)}}&=&\left[1-{C}^{F}\frac{m^{*}k_{F}^{3}}{6\pi^{2}} \right]^{2}-\widehat{W}_{1}^{(1,\pm1)}\chi_{0} \nonumber\\
&+&\left[{W}_{2}^{(1)}+C^{F} \right]\left\{ \frac{q^{2}}{2}\chi_{0}\left[1-C^{F}\frac{m^{*}k_{F}^{3}}{3\pi^{2}}\right] -2k_{F}^{2}\chi_{2}-C^{F}\frac{m^{*}k_{F}^{5}}{3\pi^{2}}(\chi_{0}-\chi_{2})\right\}\nonumber\\
&+&\left[[{W}_{2}^{(1)}+C^{F} \right]^{2}k_{F}^{4}\left\{\chi_{2}^{2}-\chi_{0}\chi_{4}+\left( \frac{m^{*}\omega}{k_{F}^{2}}\right)^{2} \chi_{0}^{2}-\frac{m^{*}}{6\pi^{2}k_{F}}q^{2}\chi_{0}\right\}\nonumber\\
&+&2\chi_{0}\left(\frac{m^{*}\omega}{q} \right)^{2}\frac{ W_{2}^{(1)}\left(1+\frac{m^{*}k_{F}^{3}}{3\pi^{2}}{X}^{(1,\pm1)}/2 \right)}{1-\frac{m^{*}k_{F}^{3}}{3\pi^{2}}\left[{W}_{2}^{(1)}-{X}^{(1,\pm1)}/2 \right]}\,.
\eqr
\ei
The coefficients $W_{i=1,2}^{(\text{S})}$ are defined as
\bqr\label{wcoeff}
\frac{1}{2}W^{(0)}_{1}&=&2(C_{0}^{\rho 0}+C_{1}^{\rho 0})+(2+\gamma)(1+\gamma)(C_{0}^{\rho \gamma_{1}}+C_{1}^{\rho \gamma})\rho_{}^{\gamma}-\left( 2C_{0}^{\Delta \rho}+2C_{1}^{\Delta \rho}+\frac{1}{2}C_{0}^{\tau} +\frac{1}{2}C_{1}^{\tau}\right)q^{2}, \nonumber\\
\frac{1}{2}W^{(1)}_{1}&=&2(C_{0}^{s,0}+C_{0}^{s\gamma}\rho_{0}^{\gamma}+C_{1}^{s,0}+C_{1}^{s\gamma}\rho_{0}^{\gamma}-\left( 2C_{0}^{\Delta s}+2C_{1}^{\Delta s}+\frac{1}{2} C_{0}^{T} +\frac{1}{2}C_{1}^{T}\right)q^{2},\nonumber\\
\frac{1}{2}W^{(0)}_{2}&=& C_{0}^{\tau} +C_{1}^{\tau}, \nonumber\\
\frac{1}{2}W^{(1)}_{2}&=&C_{0}^{T} +C_{1}^{T}.
\eqr

We have also defined the $\widehat{W}_{1}^{\text{(S,M)}}$  and $X^{(1,\text{M})}$ coefficients as
\bqr
\widehat{W}_{1}^{(0,0)}  & = & \wzu + 4 q^4 \left[ C^{\nabla J} \right]^2  
                          \frac{\pbdbt}{1+q^2 \pbdbt \left[ \wud  - C^F \right]}          \,,   \label{bbw00}   
\eqr

\bqr
\whuzu & = & - \left[ \wuu + 4 q^2 C^{\nabla s} \right]          + C^F \left[ q^2 - 4 \moqd \right]  
         +  \mrho \left[ C^F \right]^2 \left[ 2 k_F^2 + \pdemi q^2 - \frac{2}{k_F^2} \moqd \right]  \,,  \\
\whuuu & = & \wuu + 2 q^4 \left[ C^{\nabla J} \right]^2 \frac{\pbdbt}{1 + q^2 \pbdbt \left[ \wzd  \right]}
                  - 2 C^F \moqd                                                                      \nn \\
       & + & \left[ C^F \right]^2 \left( \pdemi q^2 \mrho + \tfrac{1}{16} \left[ q^2 - 4 \moqd \right]^2 \chiz 
                                                          - \pdemi k_F^2 \left[ q^2 + 4 \moqd \right] \chid
                                                          + k_F^4 \chiq \right)                  \,.     \label{bbw11}\\
\nn
\eqr
\bqr
X^{(1,0)}    & = & 2 q^2 \left[ C^F \right]^2 \frac{\pbdbt}{1+q^2 \pbdbt \left[ \wud + 3 C^F \right]}   \,, \\
X^{(1,\pm1)} & = & 2 q^2 \left[ C^F \right]^2 \frac{\pbdbt}{1+q^2 \pbdbt  \wud }          \,. \\ 
\nn
\eqr

%
%
\section{Landau approximation}
\label{app:landau}

Since we have no isospin quantum number, the $ph$ interaction is reduced to 
three terms. As done in article II we take the limit $q \rightarrow 0$ and 
$\mathbf{q}_{1,2} \rightarrow \mathbf{k}_{F}$ of the second functional derivative and  we obtain
\bqr
\label{eq:ee1a}
 V^{\text{Landau}}_{ph}(\mathbf{k}_{F},\mathbf{k}_{F}) 
& = & \pdemi {W}^{(0)}_{1,L} + \pdemi {W}^{(1)}_{1,L} \vsigma_a \cdot \vsigma_b      + \pdemi \left[ {W}^{(0)}_{2,L} + {W}^{(1)}_{2,L} \vsigma_a \cdot \vsigma_b \right] 
             2k_{F}^{2} \, \left[ 1 - \cos \theta \right]                                \nn \\
             &+&\frac{2}{3}k_{F}^{2}C^{F} \left[ 1 - \cos \theta \right]  \vsigma_a \cdot \vsigma_b +\frac{k_{F}^{2}}{3}C^{F}\frac{k_{3}}{k_{F}^{2}} S_{ab}  \,,
\eqr
where the symbols $S_{ab}$ has been defined in~\citerefsdot{Liu,Cao10}.
The various $W_{i,L}^{(S)}$ coefficients with $i=1,2$
can be easily calculated from the $W_{i}^{(S)}$ coefficients defined in Eq.~(\ref{wcoeff})
taking the limit of $W_{i}^{(S)}$ for $q\rightarrow 0$.

\noi 
So the Landau parameters can be written as
\bqr
N_0^{-1} F_0 & = & \pdemi W_{1,L}^{(0)} + k_F^2 W_{2,L}^{(0)}                           \,, \nn \\
N_0^{-1} F_1 & = &                  - k_F^2 W_{2,L}^{(0)}                            \,, \nn \\
N_0^{-1} G_0 & = & \pdemi W_{1,L}^{(1)} + k_F^2 W_{2,L}^{(1)} + \tfrac{2}{3} k_F^2 C^F  \,, \nn \\
N_0^{-1} G_1 & = &                  - k_F^2 W_{2,L}^ {(1)}+ \tfrac{2}{3} k_F^2 C^F   \,, \nn \\
N_0^{-1} H_0 & = &                                      \tfrac{1}{3} k_F^2 C^F   \,, \nn \\
\nn
\eqr
where $N_0^{-1} = \frac{2 \pi^2}{g m^*k_F}$ is the normalization factor and 
$g=2$ is the degeneracy in PNM.

\ewt
\end{appendix}



\end{document}